\documentclass[numbers]{article}
\usepackage{jcappub}
\usepackage{lmodern}
\usepackage{float}
\usepackage{graphicx}
\usepackage{amsfonts}
\usepackage{amsmath}
\usepackage{hyperref}
\usepackage{subcaption}
\usepackage{esint}
\usepackage{xcolor}
\usepackage{cancel}
\usepackage{soul}
\hypersetup{colorlinks = true, urlcolor = blue, linkcolor = blue, frenchlinks = true, pdfborder = {0 0 0}}
\setstcolor{red}
\setcitestyle{square}

\usepackage{romannum}

\begin{document}
\title{Aspects of Everpresent $\Lambda$ (\Romannum{1}): A Fluctuating Cosmological Constant from Spacetime Discreteness}
\author{Santanu Das, Arad Nasiri, and Yasaman K. Yazdi }
\affiliation{Theoretical Physics Group, Blackett Laboratory, Imperial College London, SW7 2AZ, UK}
\emailAdd{santanu.das@imperial.ac.uk, a.nasiri21@imperial.ac.uk, ykouchek@imperial.ac.uk}
\abstract{We provide a comprehensive discussion of the Everpresent $\Lambda$ cosmological model arising from fundamental principles in causal set theory and unimodular gravity. In this framework the value of the cosmological constant ($\Lambda$) fluctuates, in magnitude and in sign, over cosmic history. At each epoch, $\Lambda$ stays statistically close to the inverse square root of the spacetime volume. Since the latter is of the order of $H^2$ today, this provides a way out of the cosmological constant puzzle without fine tuning. Our discussion includes a review of what is known about the topic as well as new motivations and insights supplementing the original arguments. We also study features of a phenomenological implementation of this model, and investigate the statistics of simulations based on it. Our results show that while the observed values of $H_0$ and $\Omega_\Lambda^0$ are not typical outcomes of the model, they can be achieved through a modest number of simulations.
We also confirm some expected features of $\Lambda$ based on this model, such as the fact that it stays statistically close to the value of the total ambient energy density (be it matter or radiation dominated), and that it is likely to change sign roughly every Hubble timescale.}

\maketitle
\section{Introduction}

The interplay between UV and IR physics is at the heart of several proposed solutions to the cosmological constant puzzle  (e.g.  \cite{Cohen:1998zx, Freidel:2022ryr, Bramante:2019uub, Krotov:2010ma, Afshordi:2022ive, Perez:2018wlo, Bernard:2012nv}). This is in contrast to many of our conventional tools for describing nature, such as renormalization group theory in quantum field theory, which rely on the independence of effects at different scales. Attempts to solve the cosmological constant puzzle using our conventional tools have so far been unsuccessful, however, and point to the need for a different approach. Furthermore, the success of proposals involving a mixing of scales strongly suggests that this is the right direction to think in. This mixing of UV and IR effects also ties in nicely with the expectation that quantum gravity, the theory ultimately expected to solve the cosmological constant puzzle, will involve nonlocality in a crucial way. The challenge remains -- in the absence of a complete theory of quantum gravity -- to understand exactly \emph{how} nonlocality will relate  some large scale, such as the size of the cosmos, to the extremely small but non-zero value of the cosmological constant $\Lambda$.

The oldest example of UV-IR mixing in a cosmological context, and one that epitomises the idea, is Everpresent $\Lambda$ \cite{originallambda, Sorkin:2007bd, sorkin1994role} from causal set theory. Everpresent $\Lambda$ relies only on some basic properties of causal sets, a kind of discrete spacetime proposed to be more fundamental than the conventional continuum spacetimes we are used to. The main inputs to the Everpresent $\Lambda$ proposal are the assumptions that (1) spacetime volume plays the role of time in a path integral based dynamics, and (2) there is a statistical correspondence between the number of elements in the discrete causal set and the spacetime volume of the continuum spacetime which approximates it. From these assumptions, a simple relation can be drawn between the value of the cosmological constant at each epoch and the volume of the cosmos at that epoch, consistent with the value observed today.  Another key feature of this model is that the cosmological constant \emph{fluctuates} over the course of the evolution of the universe. We will describe all these facets of the Everpresent $\Lambda$ idea in greater detail below. 

Everpresent $\Lambda$ dates back to 1987, when Sorkin presented it as a heuristic argument \cite{originallambda, sorkin1994role}.  This was about a decade before supernova observations confirmed the value of $\Lambda\sim 10^{-121}$ in Planck units  \cite{SupernovaSearchTeam:1998fmf, SupernovaCosmologyProject:1998vns}, and many researchers at the time still anticipated its value to turn out to be zero. Already at this early stage, Everpresent $\Lambda$ predicted the magnitude of the cosmological constant to be close to its observed present value. The heuristic argument has since been fleshed out in subsequent works. For example, in \cite{ahmed2004everpresent}, the first concrete model with the basic features of Everpresent $\Lambda$ in an FLRW universe was constructed and studied via simulations; the model that we study in this paper is essentially the same as this original one. In \cite{Ahmed_2013} the incorporation of various alternative combinations of the Friedmann equations into the model in \cite{ahmed2004everpresent}  was explored; they found that similar results can be obtained under a range of combinations. In \cite{Zwane_2018}, a simpler phenomenological model (dubbed ``Model 2'') of Everpresent $\Lambda$ based on a Gaussian process, inspired by the first model but easier to simulate, was studied. In \cite{Barrow:2006vy} and \cite{Zuntz:2007sve}, observational constraints were placed on the amplitude of fluctuations, under some assumptions; these studies provide insight into what we should aim to change in future models of Everpresent $\Lambda$. Finally, in \cite{rafael_talk2}, several avenues for incorporating inhomogeneities were discussed.

Despite  progress through the works mentioned above, work on Everpresent $\Lambda$ has largely stayed close to the original heuristic argument and early models based on it. Causal set theory itself is a  minimalistic approach to quantum gravity, with causal structure and fundamental discreteness as its key inputs. Everpresent $\Lambda$ in turn, as mentioned above, uses basic properties of causal set theory as input. On the one hand, it is rather remarkable that such simple ingredients can be used to naturally explain the otherwise unnatural value of the cosmological constant. On the other hand, the simplicity of the ingredients and theoretical context are what also make it challenging to refine and build more complex models of Everpresent $\Lambda$ which can confront high precision cosmological data. Nevertheless, phenomenologically at least, we may draw on ideas outside of causal set theory to supplement and refine the Everpresent $\Lambda$ model. First, however, it is essential to better understand the model in its current state.

The aim of this work is to take steps towards placing Everpresent $\Lambda$ on firmer footing. We specifically do this by expanding on some aspects of the original heuristic argument, and by presenting a detailed study of a phenomenological model based on it. We also intend for this paper to serve as a reference point for future work on developing phenomenological models for Everpresent $\Lambda$, as well as for conducting cosmological tests against current observations. We begin in Section \ref{sec:background} with a review of the main two background topics of causal set theory and unimodular gravity. In Section \ref{sec:everpresentlambda} we present the Everpresent $\Lambda$ idea and a phenomenological implementation of it. Further explanation of our methodology is described in Section \ref{sec:method}. We carry out a detailed study of the phenomenological model presented in Section \ref{sec:everpresentlambda} and the results, including the statistical distribution of the present day Hubble parameter and the auto-correlation function of $\Omega_\Lambda$, are summarized in Section \ref{sec:results}. 
We end in Section \ref{sec:discussion} with a discussion of open questions and future directions of this program.

\section{Background Theory\label{sec:background}}
\subsection{Causal Set Theory}

Causal set theory is an approach to quantum gravity that posits that spacetime is fundamentally discrete \cite{bombelli1987space, tHooft1979, Myrheim:293594} (see also \cite{Sorkin:2003bx, Dowker:2005tz, Henson:2006kf, surya2019causal}).  This introduces a minimum length scale that naturally resolves the UV divergences of quantum field theories.  However, at macroscopic scales, this discreteness is small enough that spacetime appears to be continuous.

A causal set $\mathcal{C}$ at its most basic level is a discrete spacetime with the following properties:
     First, $\mathcal{C}$ is a partially ordered set where some elements of $\mathcal{C}$ may causally precede other elements of $\mathcal{C}$.  In particular, this intuitively means that if element $x$ precedes element $y$ ($x\prec y$), then $y$ lies in the future lightcone of $x$.
     Second, the \emph{transitivity} of the partial ordering means that if $x\prec y$ and $y\prec z$, then $x\prec z$.  In relativistic language this means that if $x\prec y$, the future lightcone of $y$ is a subset of the future lightcone of $x$.
     Furthermore, the \emph{asymmetry} of the partial ordering means that one cannot have both $x\prec y$ and $y\prec x$.  This, together with the transitivity property, implies that a spacetime cannot have closed causal loops.
     Finally, the causal set has to be \emph{locally finite}:  Between any two elements $x$ and $z$, there are a finite number of elements $y$ such that $x\prec y\prec z$.  This encodes the fundamental discreteness of spacetime in causal set theory.

The above properties describe a discrete spacetime with the basic causal structure we know from relativity.  However, the spacetime of general relativity (GR) has the further properties that it is a differentiable manifold and that it has local Lorentz symmetry, i.e., that it looks locally the same in any inertial frame.  The vast majority of causal sets with the properties enumerated in the previous paragraph, do not resemble a continuum manifold \cite{Henson:2016piq}. Therefore we need additional constraints to ensure that a causal set is manifold-like and physical.  While ideally the fundamental laws of causal set theory governing the dynamics (a work in progress) would generate spacetimes with the desired properties, in the meantime we can generate causal sets that closely approximate a continuum spacetime through a process called \emph{sprinkling}.

Sprinkling is the sampling of points from a volume of a Lorentzian spacetime manifold $\mathcal{M}$.  We would like the sampled points to satisfy the following condition:  On average, the number of sprinkled points (or elements) $N$ in a region should be proportional to the region's spacetime volume $V$. This is known as the number-volume correspondence. The proportionality factor is denoted by $\rho_{cs}=\langle N\rangle/V\equiv1/\ell_{cs}^d$, where $\ell_{cs}$ is the discreteness scale which is believed to be close to the Planck length, and $d$ is the spacetime dimension which we will take to be $4$ henceforth. In the limit $N\rightarrow\infty$ (or similarly $\ell_{cs}\rightarrow0$), we should retrieve the continuum limit. This number-volume correspondence for arbitrary volumes satisfies the requirement that the sprinkled points should preserve the local Lorentz symmetry of the spacetime \cite{bombelli2009discreteness,dowker2020symmetry}.

A sampling that satisfies the above condition is achieved through a Poisson process. In a (Poisson) sprinkled region with volume $V$, the probability of having $N$ elements is \begin{equation} \label{poisson}
    P(N)=\frac{1}{N!}e^{-\rho_{cs} V}(\rho_{cs} V)^N,
\end{equation}
i.e., the number of elements in a volume is given by the Poisson distribution.  These elements form a causal set with the relations between its elements induced by the causal structure of the underlying continuum spacetime.     As desired, the mean is $\langle N\rangle=\rho_{cs}V$, which is the statement of the number-volume correspondence, and the standard deviation $\delta N=\sqrt{\rho_{cs}V}$, which is also the smallest possible deviation \cite{saravani2014causal} from an exact number-volume correspondence.

Conversely, given a causal set $\mathcal{C}$, a Lorentzian manifold is said to be a good continuum approximation of it if $\mathcal{C}$ is a typical outcome of a Poisson sprinkling on the manifold \cite{bombelli2009discreteness}.  Lorentzian manifolds that approximate $\mathcal{C}$ (with number of elements $N$) have a volume $V=\frac{1}{\rho_{cs}}(N\pm\sqrt{N})$, i.e.,
\begin{equation}
\label{std}
    \delta V=\frac{\sqrt{N} }{\rho_{cs}}.
\end{equation}
The uncertainty arises when replacing $N$ by $\rho_{cs} V$ and vice versa in the expression for the Poisson distribution in (\ref{poisson}), and then using the known formulas for the mean and standard deviation.

So far, we have only discussed purely kinematic aspects of causal sets. Also relevant to the discussion of Everpresent $\Lambda$ is the dynamics. We could envisage a causal set growing by new  elements being born sequentially. Each new element will have some causal relations to the already existing elements. The total number of causal set elements,  which corresponds to the spacetime volume in the continuum, can thus be viewed as an external time parameter appropriate for defining the dynamics (e.g. as a constraint in a path integral). The clock ticks once every time a new element is born. Such models were studied in ~\cite{rideout1999classical,zalel2021discrete}.

On the other hand, given a causal set $\mathcal{C}$, we can define the Benincasa-Dowker-Glaser action $S_{\text{BDG}}[\mathcal{C}]$ which sums over the numbers of causal intervals of certain sizes in $\mathcal{C}$.  It has been shown that in the continuum limit, i.e. $\lim_{\rho_{cs}\rightarrow\infty}S_{\text{BDG}}[\mathcal{C}]$, this action tends to the Einstein-Hilbert action $S_{\text{EH}}$.  $S_{\text{BDG}}[\mathcal{C}]$ was first proposed in $3+1$D by \citep{benincasa2010scalar} and was later extended to other dimensions in \cite{dowker2013causal}.

In analogy with gravitational path integrals, we can use $S_{\text{BDG}}$ to define a path integral for causal sets.  Instead of integrating over different spacetime configurations (e.g. those that extend between two 3-geometries), we sum over all causal sets with a given number of elements \citep{loomis2017suppression}
\begin{equation}
\label{carlip}
    \mathcal{Z}(N)=\sum_{|\mathcal{C}|=N}e^{iS_{\text{BDG}}[\mathcal{C}]}\,.
\end{equation}
This form of the path integral (or sum over histories) for causal sets will be our motivation to introduce unimodular gravity in the next subsection.

\subsection{Unimodular Gravity} \label{subsec:ug}

Unimodular gravity was originally proposed as a mild modification of GR with the potential to solve the old cosmological constant problem, namely to address why the vacuum energy of field theories does not seem to gravitate \cite{henneaux1989cosmological, buchmuller1989gauge, weinberg1989cosmological, sorkin1997forks, BrownYork, Unruh:1988in, Unruh:1989db, sorkin1994role, Alvarez:2023utn}. As discussed in the previous subsection, holding the number of elements 
fixed is the natural constraint to impose on a path integral for causal sets (c.f. \eqref{carlip}). Motivated by this, 
let us consider the continuum gravitational path integral 
while keeping fixed the total spacetime volume $V$. In other words, to obtain the path integral with the initial and final spatial hypersurfaces $\Sigma_i$ and $\Sigma_f$, we integrate over all compact $4$-manifolds $\mathcal{M}$ that connect $\Sigma_i$ to $\Sigma_f$ and satisfy $\text{Vol}(\mathcal{M})=V$:

\begin{equation}
\label{Zglobal1}
    \mathcal{Z}_{\text{UG}}^{global}(V)=\int_{\text{Vol}(\mathcal{M})=V} \frac{\mathcal{D}g_{\mu\nu}}{\text{Vol[Diff}]}\ e^{iS[g]},
\end{equation}

where $S[g]$ is the Einstein-Hilbert action plus matter fields and vacuum energy $\tilde{\Lambda}$. The gauge group consists of diffeomorphisms that keep the total volume of a given $(\mathcal{M},g_{\mu\nu})$ unaltered. However, the total volume is an invariant under gauge transformations (in terms of a linear gauge transformation $x^\mu\rightarrow x^\mu+\xi^\mu$, $g_{\mu\nu}\rightarrow g_{\mu\nu}+\nabla_\mu\xi_\nu+\nabla_\nu\xi_\mu$, this is just the statement that $\int d^4x\sqrt{-g}\ \nabla^\mu\xi_\mu=0$, which is trivial if $\xi^\mu$ is zero on the boundary). That is why we have divided by the volume of the diffeomorphism group, Vol[Diff]. The superscript “global" refers to the fact that only the total volume of the spacetime $\mathcal{M}$ is constrained. In contrast to the global version (see \cite{sorkin1994role,PhysRevD.44.2589,Daughton:1998aa} for discussions on global unimodular gravity), there is also a local version of the theory where the volume measure $d^4x\sqrt{-g}$ is constrained to be locally equal to some (nondynamical) background 4-form $\omega$. This is the formulation of unimodular gravity that is usually discussed in the literature. It posits a seemingly stronger constraint on the metric, and reduces one of its degrees of freedom at each spacetime point. At the classical level of local unimodular gravity, it is well known (see e.g. \cite{weinberg1989cosmological, herrero2020non, Ellis:2013uxa}) that the variational principle for the constrained metric only gives the traceless Einstein equations. We shall see that the same is true for global unimodular gravity, and then conclude that at the classical level, local and global unimodular gravity are equivalent (even in their treatment of the cosmological constant).

We can implement the volume constraint in \eqref{Zglobal1} as an additional term in the action with a Lagrange multiplier $\sigma$, such that the action becomes
\begin{equation} \label{lagrangeaction}
S[g; \sigma] = S[g] - \sigma\left(\int d^4x \sqrt{-g}-V\right).
\end{equation}
To take into account this new term, we must integrate over the Lagrange multiplier such that the path integral is expressed as 
\begin{equation} \label{Zglobalnew}
    \mathcal{Z}_{\text{UG}}^{global}(V)=\int \frac{\mathcal{D}g_{\mu\nu}}{\text{Vol[Diff}]}\frac{d\sigma}{2 \pi}\ e^{iS[g]-i\sigma\left(\int d^4x \sqrt{-g}-V\right)}\,.
\end{equation}
Note that by performing the $\sigma$ integral in the path integral above, one obtains a delta function containing the volume constraint, which is how we will treat it in the next section.

Returning our attention to \eqref{lagrangeaction}, we can vary the action by varying the metric and Lagrange multiplier (after a redefinition of $\sigma$ by a factor of $8\pi G$) as
\begin{equation}
    \delta S =\frac{1}{16\pi G} \int d^4x\sqrt{-g}\ \delta g^{\mu\nu} \left(E_{\mu\nu} +\sigma g_{\mu\nu}\right) - \frac{1}{8\pi G}\delta \sigma\left(\int d^4x \sqrt{-g}-V\right),
\end{equation}
where $E_{\mu\nu}=G_{\mu\nu}-8\pi G\mkern 2mu T_{\mu\nu}+\tilde{\Lambda} g_{\mu\nu}$, and the term containing $E^{\mu\nu}$ is the variation of $S[g]$.
Setting the variations with $\delta g_{\mu\nu}$ and $\delta \sigma$ to zero independently, we obtain the equations of motion
\begin{equation} \label{einsteinlagrange}
    G_{\mu\nu}-8\pi G\mkern 2mu T_{\mu\nu}+\tilde{\Lambda} g_{\mu\nu}+\sigma g_{\mu\nu}=0,
\end{equation}
and the metric constraint equation 
\begin{equation}
    \int d^4x \sqrt{-g}=V,
\end{equation}
respectively.  Taking the trace in \eqref{einsteinlagrange}, we obtain
\begin{equation}
    -R-8\pi G\mkern 2mu T+4\tilde{\Lambda}+4\sigma=0.
\end{equation}
Solving for $\sigma$ and substituting it back into \eqref{einsteinlagrange}, we see that the $\tilde{\Lambda}$ term is cancelled, and we are left with
\begin{equation} \label{einsteintracefree}
    G_{\mu\nu}-8\pi G \mkern 2mu T_{\mu\nu}+\frac{1}{4}g_{\mu\nu}(R+8\pi G\mkern 2mu T)=0,
\end{equation}
which is the trace-free form of the Einstein equations.   The full Einstein equations can be recovered by taking the covariant derivative of \eqref{einsteintracefree}, as for example shown in \cite{weinberg1989cosmological} for local unimodular gravity.
Since $\sigma$ and $\tilde{\Lambda}$ are both constant, so must $\frac{1}{4}\left(R+8\pi G\mkern 2mu T\right)$ be, which we can therefore rename to be our effective cosmological constant $\Lambda$. We therefore have
\begin{equation}
\label{lugeom}
    G_{\mu\nu}+\Lambda g_{\mu\nu}=8\pi G\mkern 2mu T_{\mu\nu}\,.
\end{equation}
This $\Lambda$ is an integration constant, and is unrelated to any coupling constant in the action or the vacuum energy. (Note that we initially separate any vacuum energy from $T_{\mu\nu}$ and add its contribution to $\tilde{\Lambda}$, so that $T$ does not include vacuum energy). This is how unimodular gravity solves the old cosmological constant problem.  For an alternative derivation of these equations, see Appendix \ref{appendixA}.

Another way to see the same result, namely that $\Tilde{\Lambda}$ disappears from all physical quantities, is to note that the path integral with external source $J^{\mu\nu}$ is written as
\begin{equation}
\label{Zsource}
    \mathcal{Z}_{\text{UG}}^{global}(V,J^{\mu\nu})=\int_{\text{Vol}(\mathcal{M})=V} \frac{\mathcal{D}g_{\mu\nu}}{\text{Vol[Diff}]}\ e^{iS[g]+i\int d^4x\sqrt{-g}\ g_{\mu\nu}J^{\mu\nu}}.
\end{equation}
Now if we separate the vacuum energy term from the Einstein-Hilbert term and the matter action as $S=S_{\text{EH}}+S_{\text{m}}-\frac{\Tilde{\Lambda}}{8\pi G}\int d^4x\sqrt{-g}$ and use the condition $\int d^4x\sqrt{-g}=V$ of the path integral, we can simply take the non-geometric phase factor out of the integral
\begin{equation}
\label{Zsource2}
    \mathcal{Z}_{\text{UG}}^{global}(V,J^{\mu\nu})=e^{-i\Tilde{\Lambda}V/(8\pi G)}\int_{\text{Vol}(\mathcal{M})=V} \frac{\mathcal{D}g_{\mu\nu}}{\text{Vol[Diff}]}\ e^{iS_{\text{EH}}[g]+iS_{\text{m}}[g]+i\int d^4x\sqrt{-g}\ g_{\mu\nu}J^{\mu\nu}}.
\end{equation}
Physically relevant quantities, including n-point functions of the metric (if one could make sense of them in quantum gravity), are computed by functionally differentiating $\mathcal{Z}_{\text{UG}}^{global}(V,J^{\mu\nu})$ with respect to the external source and then normalizing by dividing by $\mathcal{Z}_{\text{UG}}^{global}(V,0)$ \cite{carlip2001quantum}. Therefore the factor $e^{-i\Tilde{\Lambda}V/(8\pi G)}$ cancels out, and everything is as if inside the path integral we are using the action $S_{\text{EH}+\text{m}}=S_{\text{EH}}+S_{\text{m}}$ without any vacuum energy. From now on, therefore, we will assume that the action does not include a vacuum energy term.

As is evident from the above discussion, unimodular gravity does not provide any insight on the new cosmological constant problem or on the coincidence problem. In the context of unimodular gravity, why the observed dark energy density is of the order of the present critical density remains an open problem. Everpresent $\Lambda$ brings insights from causal set theory, which together with unimodular gravity, fill this gap and provide an explanation for the observed value of the dark energy density.

\section{Everpresent $\Lambda$\label{sec:everpresentlambda}}
We now review how Everpresent $\Lambda$ follows from aspects of causal set theory and unimodular gravity presented in the previous section. A sketch of our discussion below, as well as several  alternative arguments leading to same conclusion, can be found in \cite{rafael_talk1}.

The path integral of global unimodular gravity  \eqref{Zglobal1} can be rewritten to incorporate the volume constraint using a delta function
\begin{equation}
\label{Zglobal}
    \mathcal{Z}_{\text{UG}}^{global}(V)=\int \frac{\mathcal{D}g_{\mu\nu}}{\text{Vol[Diff}]}\ \delta(V-\int d^4x\sqrt{-g})\ e^{iS_{\text{EH}+\text{m}}[g]} .
\end{equation}

The above partition function depends explicitly on $V$, the total volume of spacetime. Now instead of working in  “volume space", let us perform a Fourier transform and express the partition function in the conjugate ``$\lambda$ space'' as
\begin{equation}
\label{ft}
    \mathcal{Z}(\lambda)=\int dV\ e^{-i\lambda V}\mathcal{Z}_{\text{UG}}^{global}(V).
\end{equation}
At this point the symbol ``$\lambda$'' is an arbitrary, though suggestively chosen, symbol for the conjugate Fourier transform variable. There is no need to worry about the physical meaning of the negative values for $V$ inside the integral \eqref{ft} because \eqref{Zglobal} has a delta function at a positive value. Performing this delta function integration, and redefining $\lambda$ with a factor of $8\pi G$, we find
\begin{equation}
\label{fouri}
    \mathcal{Z}(\lambda)=\int \frac{\mathcal{D}g_{\mu\nu}}{\text{Vol[Diff}]}\ \exp\left(iS_{\text{EH}+\text{m}}[g]-\frac{2i\lambda}{16\pi G}\int d^4x\sqrt{-g}\right).
\end{equation}
The Fourier parameter $\lambda$ has turned out to play the role of the cosmological constant $\Lambda$, and $\mathcal{Z}(\lambda)$ is just the partition function of GR. The equation of motion coming from $\mathcal{Z}(\lambda)$ is exactly the same as \eqref{lugeom}, so we are justified to identify $\Lambda=\lambda$. We infer that the four-volume $V$ and the cosmological constant $\Lambda$ are quantum mechanically conjugate.
\footnote{If by conjugacy one means the existence of a commutation relation like $[\Lambda,V]\sim1$ in a canonical analysis, then there is currently no proof available. As an example of what has been done in this direction, we refer the reader to \cite{kugo2022covariant}, in which a BRST quantization of local unimodular gravity was done. In that case, the metric was expanded around some background, and the fluctuations $h_{\mu\nu}$ were quantized. Commutation relations  such as $[\lambda(p), (h^{00}+h^{33})(q)]\sim\delta(p-q)$ were obtained. However, this is still some way off from the desired commutation relations between $\lambda$ and a quantized geometric quantity such as $V$. A better understanding of quantum gravity may be needed to derive more general relations of this kind.} 
This is not a rigorous derivation, as there are many subtleties that require a full theory of quantum gravity to appreciate. However, we see it as a well motivated conjecture to expect from a quantum unimodular gravity. The uncertainty principle for these conjugate variables is
\begin{equation}
\label{uncertainty}
    \frac{\delta\Lambda}{8\pi G}\ .\ \delta V\geq\frac{\hbar}{2}
\end{equation}
Naturally one would ask what the standard deviations $\delta V$ and $\delta \Lambda$ mean. Recall from \eqref{carlip} that from the point of view of a fundamental causal set path integral, we should hold the total number of elements $N$ fixed. A consequence of this for the effective continuum picture we have been working with in this section is that we must consider spacetime volumes $V$ that are consistent with a causal set with $N$ elements. Therefore $V$ is no longer held strictly fixed as it was before, in our review of conventional unimodular gravity. We saw in the arguments leading to \eqref{std} that such volumes would be Poisson distributed with mean $\langle V\rangle=N/\rho_{cs}$ and standard deviation $\delta V=\sqrt{N}/\rho_{cs}$. We can interpret this as a wave function distribution, where the universe is in a superposition of different volumes with spread $\delta V$. Note that in allowing ourselves to work with a continuum wave function, we have implicitly assumed that we are in a regime where the number-volume correspondence makes sense. In particular, this means that we are in a macroscopic regime and not too far away from a causal set approximated by a classical geometry.

For $N, V \gg 1$ (which is indeed the case we are interested in), the Poisson distribution is very well approximated by a normal distribution $\mathcal{N}(N /\rho_{cs},\sqrt{N} /\rho_{cs})$. Moreover, if our universe is in a Gaussian wave function in $V$-space, it will also be described by a Gaussian wave function in $\Lambda$-space (through a Fourier transform). Since Gaussian wave functions saturate the uncertainty principle, we expect the state of the universe to be in a superposition of different $\Lambda$ values with spread

\begin{equation}
\label{uncertainty2}
    \frac{\delta\Lambda}{8\pi G} =\frac{\hbar}{2} \frac{1}{\delta V}\,.
\end{equation}
Finally, using \eqref{std} and the relation between number and volume, we find
\begin{equation}
\label{Heis}
    \delta\Lambda=4\pi\left(\frac{\ell_p}{\ell_{cs}}\right) ^2 \frac{1}{\sqrt{V}}\,.
\end{equation}
We used $\ell_p=\sqrt{G\hbar}$ as the Planck length and $\ell_{cs}=\rho_{cs}^{-1/4}$ for the discreteness scale (in $4$ dimensions) of causal sets. 

The standard deviation of $\Lambda$ in \eqref{Heis} can be estimated for our universe by assuming the discreteness length to be of order of the Planck length, and
 setting $V$ to the volume of the observable universe. 
 
 By dimensional analysis,  in an FLRW universe this volume should be of order $H^{-4}$ in fundamental units, $H$ being the Hubble parameter. In our background solutions for the Friedmann equations in the next section, we will see that this is indeed the case. Roughly then,
    \begin{equation}
        \delta \Lambda\sim H^2.
    \end{equation}

We will assume that the mean value about which $\Lambda$ fluctuates is zero.\footnote{This is a consistent choice since the mean of $\Lambda$ is arbitrary (or at least undetermined) in unimodular gravity. It remains the case even if one takes into account the vacuum energy $\Tilde{\Lambda}$ (from the cosmological constant term in $S_{\text{EH}}$).  To see this, we can absorb $\Tilde{\Lambda}$ into the definition of the Fourier variable $\lambda$ in \eqref{fouri}.  This amounts to a constant shift in the effective cosmological constant $\Lambda=\Tilde{\Lambda}+\langle\lambda\rangle+\delta\lambda$.  But since $\langle\lambda\rangle$ is arbitrary, we can choose it so that the shifted mean $\Tilde{\Lambda}+\langle\lambda\rangle$ is 0.  A more complete theory should advise the correct mean for $\Lambda$.  The fluctuations $\delta\lambda$ remain unaffected by the vacuum energy.}
This assumption does not stem from any fundamental aspect of causal set theory or unimodular gravity and will require justification in the future. See \cite{hawking1984cosmological, coleman, Ng:1990rw} for some possible arguments in favor of this, using ideas from Euclidean quantum gravity. See also footnote $4$ in \cite{ahmed2004everpresent} and towards the end of \cite{rafael_talk2} for a Lorentzian guess (involving a non-manifoldlike phase of the universe) at how this may come about. Therefore, $\Lambda=\langle\Lambda\rangle\pm \delta \Lambda\sim H^2$, which is indeed a correct estimate for the current value of $\Lambda$. Furthermore, this is telling us that $\Lambda$ is not temporally constant in our picture. It fluctuates such that at any time in the expansion history of the universe, its standard deviation is of the order of $1/\sqrt{V}$ (in FLRW this is comparable to the critical energy density at that time); hence the name \emph{Everpresent} $\Lambda$, as it will always be relatable to the physical scale of the universe in this way.

\subsection{A Phenomenological Implementation - Model 1}

Perhaps the closest analog to the $\delta\Lambda\mkern 1mu\delta V$ uncertainty relation is the energy-time uncertainty in quantum mechanics. It has been suggested \cite{sorkin1994role,gielen2022quantum, Gielen:2013naa} that in quantum gravity, an increasing spacetime volume $V$ (such as the volume of the past lightcone of an observer) could serve as a \textit{time} variable to treat the seemingly timeless nature of GR and the Wheeler-DeWitt equation.

Just as the $\delta{E}\mkern 1mu\delta t$ uncertainty in quantum mechanics alone cannot provide a dynamical model for energy non-conservation in short time intervals, the $\delta\Lambda\mkern 1mu\delta V$ uncertainty relation does not provide a practical model for how to evolve $\Lambda$. In \cite{ahmed2004everpresent} the authors suggested a phenomenological model (referred to as ``Model 1'' in \cite{Zwane_2018}) for the dynamics of $\Lambda$ that shows the expected fluctuations in \eqref{Heis}. Their strategy was to build a classically stochastic model for the fluctuations in $\Lambda$, working with a single universe rather than a superposition.  We will now review this model whose properties will be our focus for the remainder of this paper.

From the causal set point of view, and due to the number-volume correspondence, \eqref{Heis} is equivalent to
\begin{equation}
\label{Heis2}
    \delta\Lambda=\frac{8\pi\alpha}{\ell_{cs}^2} \frac{1}{\sqrt{N}},
\end{equation}
where we have defined
\begin{equation} \label{alpha}
    \alpha\equiv\frac{1}{2}\left(\frac{\ell_p}{\ell_{cs}}\right)^2.
\end{equation}
We next make the ansatz that \eqref{Heis2} is true even element-wise, in the sense that each causal set element contributes a random variable with standard deviation $8\pi\alpha/\ell_{cs}^2$ to $\Lambda$. The simplest such scenario would be to think of the single element contribution to $\Lambda$ as a Bernoulli random variable:
\begin{equation}
    \Lambda|_{\text{1 element}}=\Lambda|_x=\begin{cases}
    \phantom{-}\frac{8\pi\alpha}{\ell_{cs}^2},\ \ \ p=\frac{1}{2}\\
    -\frac{8\pi\alpha}{\ell_{cs}^2},\ \ \ p=\frac{1}{2}
    \end{cases}.
\end{equation}
(In fact any other distribution with mean $0$ and standard deviation $8\pi\alpha/\ell_{cs}^2$ would work just as well for our purposes). 
The cosmological constant part of the action is obtained by summing over all the above stochastic variables in the $N$-element (sub)causal set corresponding to the region with volume $V$: 
\begin{equation} \label{discsum}
    S_\Lambda=\frac{\ell_{cs}^4}{8\pi\ell_p^2}\sum_{x\in C_V}\Lambda|_x\,.
\end{equation}

We have replaced the integral $\int d^4x\sqrt{-g}$ with the discrete sum $\ell_{cs}^4\sum_{x}$.  Also, a minus sign has been absorbed into the definition of the cosmological constant. Whatever the distribution of $\Lambda|_x$ is, if its standard deviation is $8\pi\alpha/\ell_{cs}^2$, $S_\Lambda$ will follow a normal distribution $\mathcal{N}(0,\sqrt{N}/2)$ according to the central limit theorem (recall that $N, V \gg 1$). 

As already mentioned, by summing over the elements $x$ corresponding to a given region of spacetime  in \eqref{discsum}, we obtain an effective $\Lambda$ for the region. So far we have not discussed how to choose this region.  We will follow \cite{ahmed2004everpresent} and choose it at any time of interest to be the volume $V$ of the past lightcone of a point $x$ at that time. By making this choice, not only is $V$ causally defined, it is also conveniently parametrized by the point as $V(x)$. In the same way we can also express the effective $\Lambda$'s volume-dependence as a function $\Lambda(x)$ (not to be confused with the element-wise $\Lambda|_x$).

We interpret $\Lambda(x)$ that appears in \eqref{fouri} then as the effective cosmological constant that is ascribed to the past lightcone of the point $x$ such that 

\begin{equation} \label{Slam2}
    S_\Lambda|_{\text{past of $x$}}=\frac{1}{8\pi G}\Lambda (x) V(x)= \frac{\ell_{cs}^2}{16\pi\alpha}\Lambda(x)N(x)\,.
\end{equation}
$V(x)$ is the volume of the past lightcone at $x$, and $N(x)$ is the corresponding number of causal set elements. Therefore, the distribution of $\Lambda(x)$ turns out to be 
\begin{equation}
\mathcal{N}(0,\frac{8\pi\alpha}{\sqrt{V}}).
\end{equation}
Thus we have provided a stochastic dynamics that satisfies \eqref{Heis}, as desired. 

Now we can obtain a recursive formula for $\Lambda(x_n)$. Imagine we have found $S_\Lambda|_{\text{past of $x_{n-1}$}}$ and $\Lambda(x_{n-1})$ for some point $x_{n-1}$ in the past of $x_n$, close enough to $x_n$ to capture the variations in $\Lambda(x)$. The additional contribution to $S_\Lambda|_{\text{past of $x_n$}}$ compared to $S_\Lambda|_{\text{past of $x_{n-1}$}}$ is from all random variables $\Lambda|_z$ such that $z$ is in the past of $x_n$ but not in the past of $x_{n-1}$. The number of such points is proportional to the volume they occupy, $\Delta V_n= V(x_n)-V(x_{n-1})$, and the sum of their contributions to $S_\Lambda$ is, again from the central limit theorem, a random variable $\sim\sqrt{\Delta V_n}/(2 \ell_{cs}^2)\mathcal{N}(0,1)$. Therefore if we generate a standard normal random variable $\xi_n$, we can write
\begin{align}
    \Lambda(x_n)&=\frac{8\pi\ell_p^2 S_\Lambda|_{\text{past of $x_{n-1}$}}+8\pi\alpha\sqrt{\Delta V_n}\,\xi_n}{V(x_n)}\\&=\frac{\Lambda(x_{n-1})V(x_{n-1})+8\pi\alpha\sqrt{\Delta V_n}\, \xi_n}{V(x_n)}.
\end{align}

Summing this recursive relation back to some initial point $x_0$ in the far past of $x_n$, we find
\begin{equation}
\label{stoch}
    \Lambda(x_n)=\ \frac{8\pi G S_0+8\pi\alpha\sum_{i=1}^{n}\sqrt{\Delta V_i}\,\xi_i}{V(x_n)},
\end{equation}
where $8\pi GS_0$ is $\Lambda(x_{0})V(x_{0})$, or possibly an initial condition coming from a pre-geometric regime (see Section~\ref{subsec:ic}).

In the above derivation we have explicitly kept track of $\ell_p$ and $\ell_{cs}$, such that we can interpret  the parameter $\alpha$ that sets the magnitude of the fluctuations as the square of the ratio of the Planck length to the discreteness length, as in \eqref{alpha}.\footnote{In \cite{ahmed2004everpresent, Ahmed_2013, Zwane_2018} the scale of the fluctuations was more loosely defined, i.e. not in terms of any explicit expression connected to other scales in the problem.  This is perhaps to allow for more freedom in the parameter, coming from relaxing some of the assumptions.} We should only take this interpretation with a grain of salt, however, as we started with the assumption that each causal set element contributes a random value to the total $\Lambda$, while in fact the true causal set mechanism that gives rise to Everpresent $\Lambda$ is not fully understood. Nonetheless, our account of $\alpha$ as the ratio of scales is consistent with the predictions from the path integral that led to \eqref{uncertainty}.  For example, we see that if we take the continuum limit and send $\ell_{cs}\rightarrow 0$, then $\alpha \rightarrow \infty$.  Since in the continuum we have definite volumes with no uncertainty, this limit corresponds to $\delta V \rightarrow 0$, and therefore from \eqref{uncertainty2} we expect infinite fluctuations in the cosmological constant.  Thinking of $\alpha$ in this way is also useful in that it guides the kinds of values we would expect it to take, since we believe the discreteness scale to be close to the Planck scale.

\section{Methodology\label{sec:method}}

Before we study some characteristic features emerging from simulations of Everpresent $\Lambda$  based on Model 1,  we must first say a few more words regarding our methodology and implementation of the model. This section will cover these details.
\subsection{An Isotropic and Homogeneous Universe}\label{subsec:frw} 
In the previous section we ended by providing a stochastic model that describes the evolution of $\Lambda$ over discrete time steps. This evolution was expressed as a function of a random parameter and the volume $V$ of a point's past lightcone at those instances.  Next we need to know the metric with which to calculate $V$, and how this metric evolves as a result of $\Lambda$. 

We will follow \cite{ahmed2004everpresent} by assuming we have a spatially flat, isotropic and homogeneous cosmology, represented by the Friedmann-Lema\^itre-Robertson-Walker metric (in conformal time coordinates)
\begin{equation} \label{frw}
    ds^2=a(\tau)^2(-d\tau^2+dr^2+r^2d\Omega^2).
\end{equation}
In standard GR the evolution of the scale factor $a$ is described by the Friedmann equations:
\begin{equation}
     H^2= \left(  \frac{   \dot{a}}{a}\right)^2=\frac{8\pi G}{3}\left(\rho_{m}+\rho_{r}\right)+\frac\Lambda 3
    \label{Friedman1},
\end{equation}

\begin{equation}
\frac{ 2 \ddot{a}}{a}+\left(  \frac{   \dot{a}}{a}\right)^2=-8\pi G(p_{r}+p_\Lambda)
\label{Friedman2},
\end{equation}
where the derivatives are with respect to the cosmic time $t$, related to the conformal time $\tau$ in \eqref{frw} by $dt=a \, d\tau$.

Any implementation of Everpresent $\Lambda$ represents a departure from classical GR, because the Einstein equation does not admit a varying cosmological constant (see Appendix \ref{subsec:fluid} for a discussion on this). Our strategy (similar to what was done in \cite{ahmed2004everpresent}) will be to use only   \eqref{Friedman1} to generate our Everpresent $\Lambda$ histories.

We can obtain $V$ by integrating the volume element coming from \eqref{frw}, $dV=a^4r^2\sin\theta\ d\tau \mkern 2mu dr\mkern 2mu d\theta \mkern 2mu d\phi$, over the past lightcone of a given point $x$ at conformal time $\tau$, to the future of some initial time $\tau_i$. The 3d integration of the volume element at a fixed time slice $\tau'$ gives $d\tau'\ \frac{4\pi}{3}a^4r(\tau')^3$, where $r(\tau')=\tau-\tau'$ is the radius of the past lightcone on that hypersurface. Therefore the $\tau'$-integral remains to be calculated:\footnote{The expression in \eqref{vlightcone} can be seen to be equivalent to the integral 
$
    V(t)=\frac{4\pi}{3}\int_{t_i}^{t}dt'\ a(t')^3\left(\int_{t'}^{t} dt''\ \frac{1}{a(t'')}\right)^3
$ in \cite{ahmed2004everpresent} through the coordinate transformation $d\tau = \frac{dt}{a}$. }
\begin{align}\label{vlightcone}
V(\tau)&=\frac{4 \pi}{3} \int_{\tau_i}^{\tau} d \tau^{\prime} a(\tau^{\prime})^{4}( \tau - \tau' )^{3} \\ \label{vint}
&= \frac{4 \pi}{3} \left[ \tau^3 \int_{\tau_i}^{\tau} d \tau^{\prime} a(\tau^{\prime})^{4}
- 3 \tau^2 \int_{\tau_i}^{\tau} d \tau^{\prime} a(\tau^{\prime})^{4}\tau^{\prime}
+ 3 \tau \int_{\tau_i}^{\tau} d \tau^{\prime} a(\tau^{\prime})^{4}\tau^{\prime 2} 
- \int_{\tau_i}^{\tau} d \tau^{\prime} a(\tau^{\prime})^{4}\tau^{\prime 3}\right].
\end{align}

\subsection{Numerical Recipe}\label{subsec:numrec}

We now describe our numerical procedure for evolving $\Lambda$ and $V$.  Although we have thus far described quantities as evolving with time, we will find it useful to instead use the scale factor $a$ as our ``time'' parameter,\footnote{This is because the Friedmann equation \eqref{Friedman1} is already in terms of $a$, and in practice the simulations also tend to run more quickly when working with $a$.} and calculate the evolution of the conformal time $\tau$. We divide $a$ into $n$ equal parts so that at step $i\ge i_{initial}$ of our simulation, the value of $a$ is $i/n$.  In this picture $a=0$ represents the time of the big bang, and $a=1$ the present day. The convention to use equal size steps in $a$ is a choice that gives reasonable results in our analysis.  However, one may experiment in varying the step size, e.g. step sizes that start out small and gradually become larger.

Before we can start a simulation, we must choose our initial parameters.  These parameters are $S_0$, $V_0$, the number of time steps $n$, and the time step from which to start the evolution $i_{initial}$. Since we do not yet know the details of the quantum gravitational physics that preceded the Everpresent $\Lambda$ era, we must make some assumptions in these choices. See Section \ref{subsec:ic} for further discussion on this. We also choose the matter and radiation densities $\rho_{m}$ and $\rho_{r}$ so that they are consistent with their present day values in the expression for $d\tau/da$ below. We also have to pick a seed number for our random number generator. Any seed number contains in itself a unique sequence of Gaussian random numbers $\xi_i$.

At each time step $i$ in our simulation, each of our variables update as follows:
\begin{enumerate}
    \item Increment $a_{i-1}\rightarrow a_i$.
    \item Using $\Lambda_{i-1}$ from the previous time step, we calculate $d\tau/da$ as a function of (arbitrary) $a$ as 
    \begin{align}\label{sqrteninv0}
        \frac{d\tau}{da} &= \frac{1}{a\dot{a}} = \sqrt{\frac{1}{a^4 H_i^2}}\\ \label{sqrteninv}
  &= \sqrt{\frac{3}{8\pi G  a^4\left( \rho_{m} + \rho_{r}\right) + a^4 \Lambda_{i-1}}}\,.
    \end{align}
    \item Given $a_i$, we calculate $V_i$ by  changing the integration variable in  \eqref{vint} to $a$ and computing the sum
    \begin{equation}
    V_i = \sum_{j=0}^3 c_j\tau(a_i)^j V_{j, i}\,,
    \end{equation}
    where
    \begin{equation} \label{vji}
    V_{j, i} = V_{j, i-1}+\int_{a_{i-1}}^{a_i}da \, \tau^{3-j} \frac{d\tau}{da}a^4,
    \end{equation}
      i.e.,  we add the change in the four integrals of \eqref{vint} in the interval $[a_{i-1}, a_{i}]$ to their previous values, where $c_j$ are the appropriate prefactors. The time function $\tau(a)$ is itself defined as the integral
    \begin{equation} \label{etaint}
    \tau(a) = \tau(a_{i-1}) + \int_{a_{i-1}}^a da \frac{d\tau}{da},
    \end{equation}
   where $\tau(a_{i-1})$ is saved from the previous time step.
    \item Next, we generate a random number $\xi_i$ (which is determined by our choice of the seed number) and update the action from the numerator of \eqref{stoch} to obtain
    \begin{equation}
    S_i = S_{i-1}+\alpha\sqrt{V_i - V_{i-1}}\, \xi_i/G,
    \end{equation}
    and update $\Lambda$ as 
    \begin{equation} \label{lambdai}
        \Lambda_i = 8\pi GS_i/V_i.
    \end{equation}

\end{enumerate}

The integrals in \eqref{vji} and \eqref{etaint} are numerically evaluated using Romberg's method, to speed up the convergence.

\begin{figure}
         \centering
         \includegraphics[width=.8\textwidth,trim = 0 250 0 270,clip]{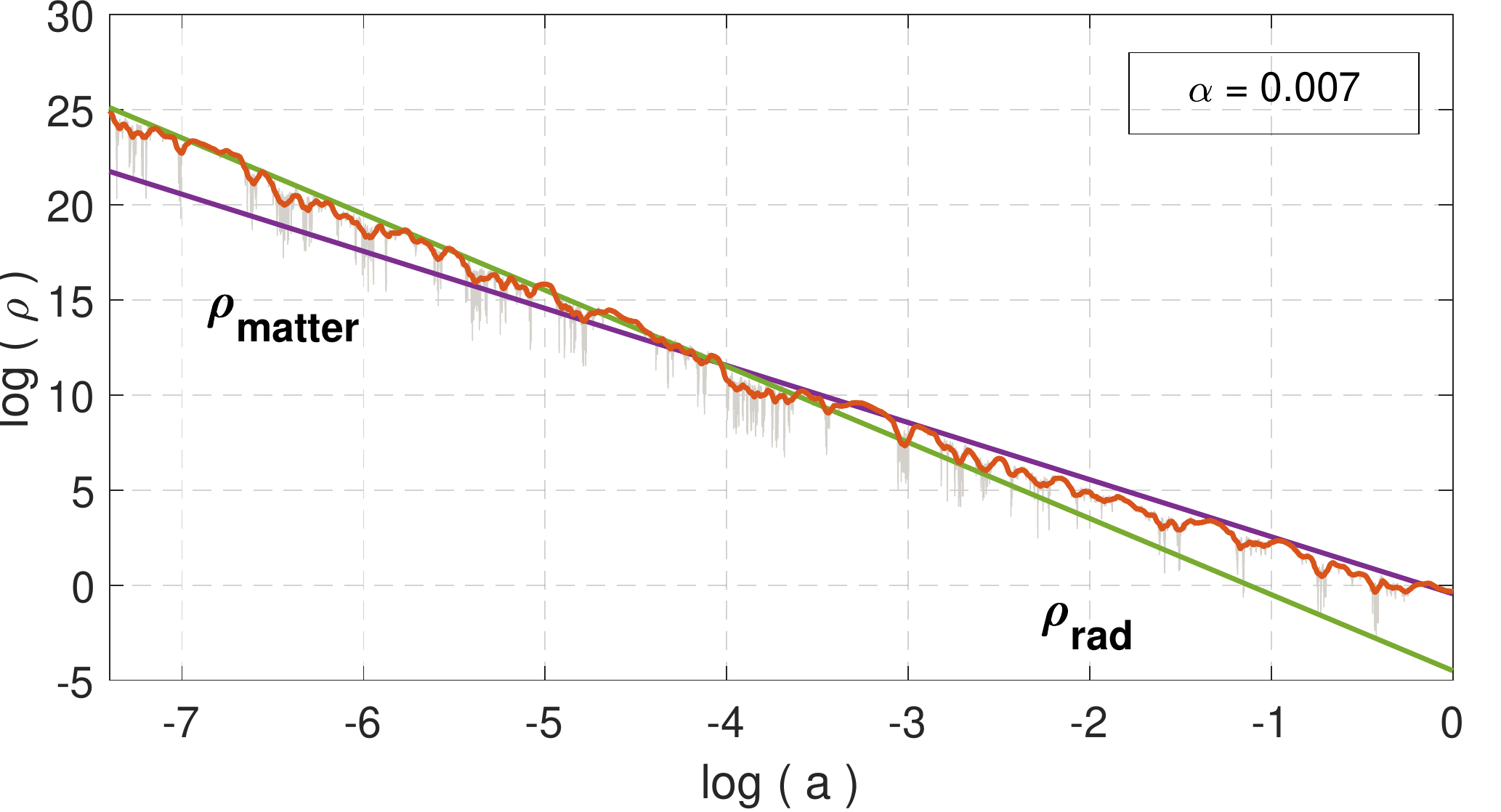}
         \caption{The evolution of matter (purple) and radiation (green) density in the universe is shown. The thin gray curve is the absolute value of the energy density of (Everpresent $\Lambda$) dark energy for a particular realization of our simulation. A smoothed-out version of the dark energy density is shown in red. The plot shows that the dark energy density is always of the order of the ambient density, be it radiation (early on) or matter (later). All the densities are normalized by the critical density, such that the present-time total energy density is $1$. Here the dimensionless model parameter $\alpha$ is set to 0.007. }
         \label{fig1}
     \end{figure}

Figure \ref{fig1} shows an example profile for a fluctuating $\Lambda$ produced in this way.  As can be seen in this figure, the magnitude of $\rho_\Lambda$ tracks the ambient energy density when following the above implementation, as expected from $|\Lambda|\sim 1/ \sqrt{V}\sim H^2$.

It is also an implicit assumption that $a$ must always have a positive time derivative, even though plausibly we could end up with a collapsing universe.  However, given that our universe is expanding, this is not an important detail. 

Having all the necessary equations at hand, we can perform simulations of the background cosmology. The statements we can make are no longer deterministic. We should instead look at the final distribution of various physical quantities after performing many realizations of the stochastic process. 

\subsection{$H^2<0$ Crashed Runs} \label{subsec:h2lt0}
It is important to note that since there are no additional constraints on the random numbers $\xi_i$, it is possible for the fluctuations of $\Lambda$ in \eqref{stoch}  to cause the total energy density  $\rho_{tot}$, and therefore also $H^2$, to become negative.  When this happens $d\tau/da$  becomes imaginary in \eqref{sqrteninv}, and the spacetime volume a complex number, which does not make sense classically.

If one takes Model $1$ seriously, then it should be regarded as a legitimate possibility that $H^2$ can become negative.  It would not be correct to view this occurrence as a sign of unphysical input parameters. This is because, statistically speaking an evolution will almost surely reach a point where the random fluctuations in $\Lambda$ will cause $H^2$ to become negative, and it is only a matter of how long the evolution according to this model lasts before this happens.

A negative $H^2$ and $\rho_{tot}$ is a sign that our effective picture is no longer valid, and is due to the use of the Friedmann equation in \eqref{sqrteninv0} and \eqref{sqrteninv}.  The Friedmann equation must be replaced by some other (as yet unknown) dynamics at this point, and one can speculate what the evolution of the universe will look like when this occurs.  It may be that a different effective but still classical evolution equation should be used under such circumstances. Alternatively, a quantum dynamics may be needed.

As the dynamics in the $H^2<0$ regime is unknown at present, we will disregard universes that reach $H^2<0$ before the present, and terminate their evolutions at this point.  When this happens, we refer to it as an $H^2<0$ \emph{crash}, as the number in the square root of \eqref{sqrteninv} becomes negative, and would cause a computer code requiring real variables to crash. We again emphasize  the \emph{physical} origin of these crashes, as they cannot be avoided by fixing a programming error.  

There is a higher likelihood for an $H^2<0$ crash to happen at earlier times, for larger values of $\alpha$, and when the simulations are performed over longer time scales. The fact that the $H^2 < 0$ crashing frequency becomes greater the earlier the start time of the simulations is, seems to suggest that there might be a link between these crashes and the breakdown of the model's semi-classical approximations. These $H^2<0$ crashes are an aspect of Model 1 that needs to be understood better and we devote Section \ref{subsec:crashes} below to further discuss some aspects of these simulations. See also Section \ref{subsec:ic}.

\subsection{$\rho_m$ as a Scaling Parameter at Late Times}\label{subsec:rhom}

Model 1 shows a very curious scaling invariance as an artifact of treating the seed number as a parameter. For a fixed seed number $s$, although $(\Omega_m^0h^2,\alpha)$ are independent parameters of the model, $(\Omega_m,\alpha)$ are perfectly correlated in the matter-dominated era. This is because if we keep the seed and $\alpha$ fixed, and then make the change 
\begin{equation}
\rho_m^{0}\rightarrow c_0\rho_m^{0},
\end{equation}
the quantities $\Lambda$, $H^2$, and $1/\sqrt{V}$ also get re-scaled by $c_0$ (assuming that $S_0=0$), as can be seen from \eqref{sqrteninv0} - \eqref{lambdai}. In particular, this means that the ratio $\rho_m/\Lambda$ and hence $\Omega_m$ and $\Omega_\Lambda$ remain unchanged for fixed $\alpha$ and random number series $\{\xi_i\}$ in \eqref{stoch}. In addition, $\Omega_\Lambda$ remains unchanged after a re-scaling of $\rho_m$ in the radiation domination era as well, but this time simply because $\rho_m$ is sub-dominant in the background. Therefore, for a fixed seed, $\Omega_m$ is only a function of $\alpha$ and not of $\rho_m$; $(\Omega_m,\alpha)$ are perfectly correlated. 

Using the above fact, there is a simple way to reconstruct the expansion history in the matter-dominated era for $(s,\Omega_m^0h^2,\alpha)$ if we know the expansion history for $(s,0.14,\alpha)$, where 0.14 is just a fiducial choice for the matter density. The latter gives us the knowledge of $\Omega_\Lambda$ at different redshifts. Together with $\rho_m(a)$, this is enough to find $H$ and hence the expansion history, without having to perform numerically expensive volume integrals. In addition, if the realization crashes due to encountering a negative $H^2$ for some $(s_1,0.14,\alpha_1)$, then the same would happen for every $(s_1,\Omega_m^0h^2,\alpha_1)$, because of the scaling property of $\Omega_m^0h^2$, and also for every $(s_1,0.14,\alpha_2)$ with $\alpha_2$ slightly bigger than $\alpha_1$. Intuitively, this can be understood by remembering from \eqref{stoch} that $\alpha$ controls the magnitude of the fluctuations in $\Lambda$, although not in a linear way since $V$ in the denominator also depends on $\alpha$. Nevertheless, an $H^2<0$ crash means that $\rho_\Lambda$ has become equal to $-\rho_m$ at some time, and now a slight increase in $\alpha$ makes $\rho_\Lambda$ more negative. The  $H^2<0$ crash will therefore occur in this case as well. This was confirmed by our numerical simulations. Hence, $\alpha_1$ is an upper bound on $\alpha$ to avoid  $H^2<0$ crashes, in the fixed seed $s_1$.\footnote{This $\alpha_1$ is not a global bound, however. As explained in Section \ref{subsec:ic}, for some seeds  there is another regime of large values of $\alpha$, beyond their corresponding $\alpha_1$, for which the simulations do not crash. This regime, however, is not of our interest in the present paper, as it describes expansion histories that are strongly dominated by dark energy.}

\section{Results and Further Investigations\label{sec:results}}
 We summarize below some results highlighting both qualitative properties and quantitative statistics of Model 1 simulations of Everpresent $\Lambda$.
\subsection{Statistical Distributions of Simulation Outcomes}\label{subsec:stat}
We carry out a set of simulations to understand the statistical properties of Model 1. In Model 1, the free background parameters are $(\Omega_m^0 h^2, \alpha)$. For our universe $\Omega_m^0 h^2 \approx 0.14$, and as this parameter corresponds to the matter density of the universe (which is a measurable quantity), a significant variation in it is not possible (see \cite{das2023aspects} for an analysis related to this). Therefore, in this subsection we fix $\Omega_m^0 h^2=0.14$ and vary $\alpha$ in order to study the properties of Model 1.  We report the statistics from 3 choices of values for $\alpha$: $0.005$, $0.01$ and $0.012$. For each of these $\alpha$'s we run $10,000$ simulations with different seeds and subsequently study the runs that reached the present day (i.e. those that did not encounter an $H^2<0$ crash).  For $\alpha=0.012$, only $535/10000$ runs successfully completed.  We did not consider larger values of $\alpha$ in this subsection's analysis, as we would have lacked a sufficient number of completed runs.
\begin{figure}[H]
    \centering
    \begin{subfigure}[h]{0.95\textwidth}
         \centering
         \includegraphics[width=\textwidth]{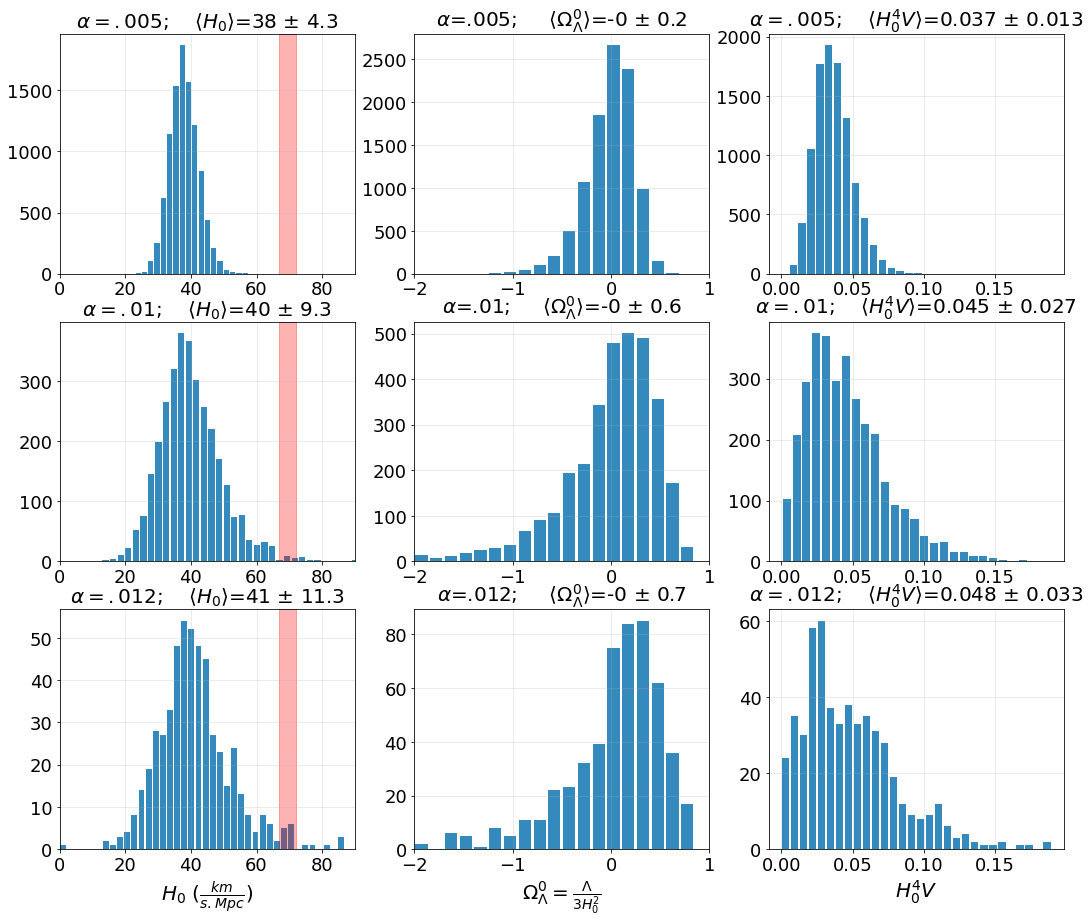}
     \end{subfigure}
    \caption{The plot shows the distribution of $H_0$, $\Omega_\Lambda^0$, and $H_0^4V$ (where $V$ is the present day volume) for Model 1 with $\Omega_m^0h^2=0.14$. The three rows show the plots for $\alpha=0.005,0.01,$ and $0.012$ respectively. The interval $H_0\in[67,72]{\rm km/s/Mpc}$ is shaded in red in the first column to show the presently accepted range of the Hubble parameter. As $\alpha$ increases, more runs terminate because of reaching negative values of $H^2$. The number of completed runs for the three $\alpha$'s are 9990, 3228, and 535 respectively each out of 10000.}
\label{fig3:model1_stat}
\end{figure}
The results from our analysis are shown in Figure \ref{fig3:model1_stat}.  The first column shows the distribution of $H_0$ (the Hubble parameter at present time). For $\alpha=0.005$, around $99.9\%$ of the simulations completed the runs. However, none of them come close to the observed value of $H_0$  of $72\, \text{km/s/MPc}$. The peak of the distribution is about 7.9 standard deviations away from this value. For $\alpha=0.01$ and  $0.012$, the peaks are about 3.44 and 2.74 standard deviations away, respectively, and we see that a few data points give reasonable values of $H_0$. In the figure, we have put a red band indicating the roughly accepted range $H_0 = [67, 72]\, \text{km/s/MPc}$. 

It is interesting to note that the mean of $H_0$ in all three cases is close to $40\, \text{km/s/MPc}$. This is because, at any time, $\Lambda$ is a  Gaussian random number with a mean of zero. The variance at each time will vary depending on the 4-volume of the past lightcone at that time. Provided  the runs all succeed, we can expect the Hubble parameter to peak at the value where the universe has no dark energy. If we had a universe filled with only matter and radiation, the present-time Hubble parameter would be around $37.7\text{km/s/MPc}$. As we are excluding the runs that crashed  due to encountering negative values of $H^2$ before reaching today, we expect the peak of the distribution to shift to higher values. We can see this trend from our distributions. As $\alpha$ increases, the number of rejected samples increases, shifting the peak of the Hubble parameter further to the right.

The second column shows $\Omega_\Lambda^0=\rho^0_\Lambda / (\rho^0_\Lambda + \rho^0_m)$. 
The value of $\Omega_\Lambda$ must at all times be less than $1$ and have mean zero, and both of these features can be seen in the plots.  With increasing $\alpha$, the standard deviation of the distribution is also increasing. 

The final column shows the distribution of $H_0^4 V$, where $V$ is the present day volume. We expect $V$ to be proportional to $H_0^{-4}$. Here also, the standard deviation of the distribution increases with increasing $\alpha$. The peak of the distribution again roughly corresponds to a zero dark energy universe, for which $H_0^4V = 0.033$.

\subsection{Frequency of Completed Runs}\label{subsec:crashes}

As a stochastic model, Model 1 produces a unique expansion history for every seed (which in turn gives a series of random numbers) and $\alpha$. As mentioned in Section \ref{subsec:h2lt0}, in many cases we obtain a negative total energy density before reaching the present. This leads to an imaginary Hubble parameter, which in turn would result in a complex spacetime volume if the steps described in Section \ref{subsec:numrec} are followed.  We therefore discard these runs as the correct dynamics beyond this point is not known. There are a number of factors which  affect the likelihood to encounter an $H^2<0$ crash. One is the value of $\alpha$, which controls the magnitude of fluctuations of the dark energy density. Increasing $\alpha$ raises the magnitude of fluctuations, increasing the probability of a particular run to crash. The choice of time step during the expansion history calculation is another critical factor. The dark energy density remains constant between every two consecutive time steps and it updates at the end of every time step. Even though ideally we would like the time step to be of the order of the discreteness scale, it is not computationally feasible to implement it that way. Instead we choose a more modest time step in terms of the scale factor ($a$), time ($t$), conformal time ($\tau$), or some other suitable quantity. Decreasing the value of the time step increases the number of them and with it the number of times the dark energy density fluctuates. Furthermore, increasing the number of fluctuations increases the chances for positive and negative contributions to cancel one another, decreasing the possibility of an $H^2<0$ crash. 

\begin{figure}[H]
    \centering
    \includegraphics[width=.7\textwidth,trim = 0 250 0 250,clip]{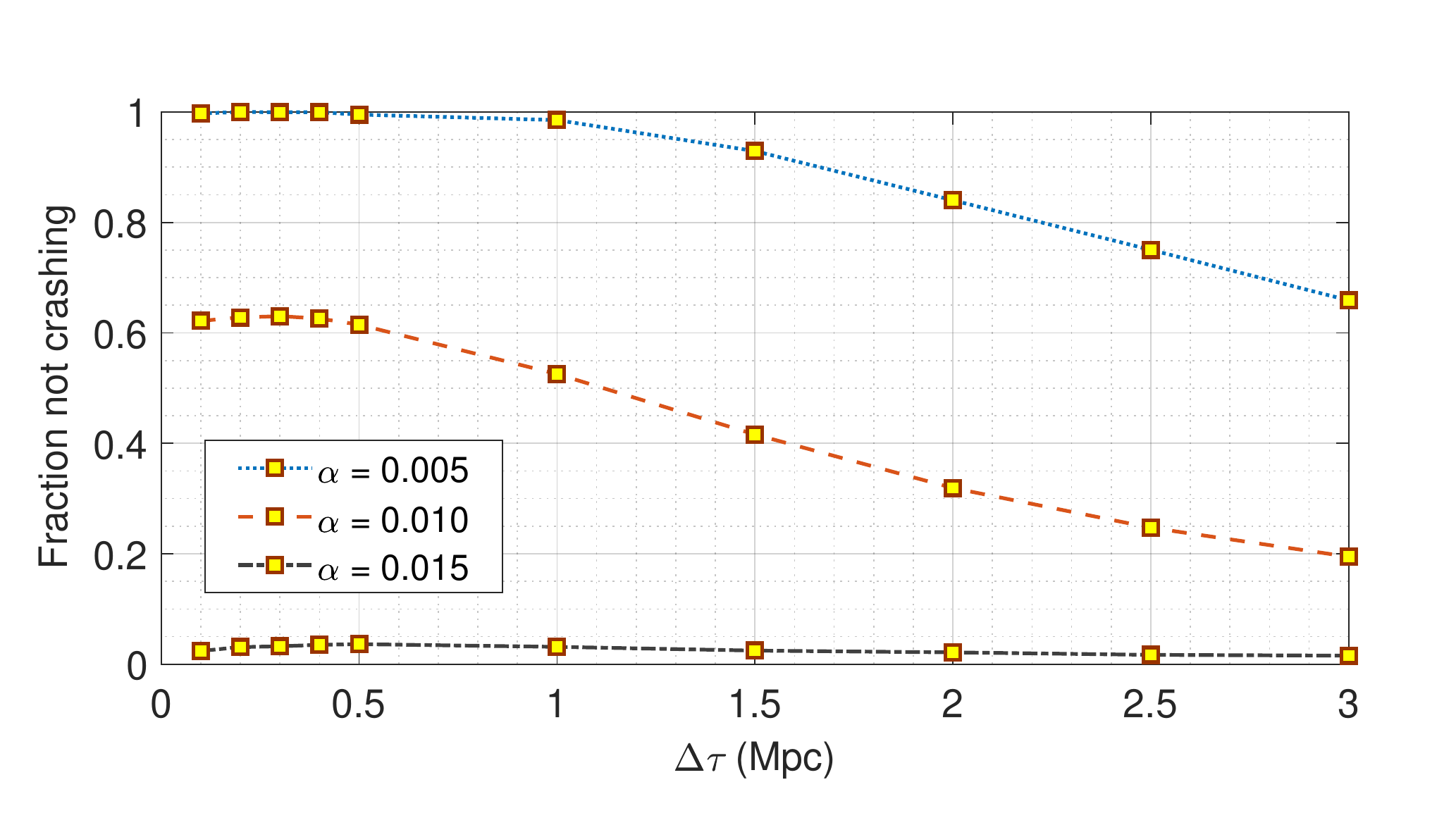}
    \caption{The plot shows how the time step affects the crashing fraction for $\alpha =$ 0.005, 0.001, and 0.015. As $\alpha$ increases, the fraction of completed runs drops considerably. For higher $\alpha$ the crashing fraction almost tends to 100\%. As we decrease the time step, $\Delta\tau$, the crashing fraction decreases, eventually saturating to a fixed value. }
    \label{fig:crashinganalysis}
\end{figure}

We run a set of simulations to confirm the observations of the previous paragraph and further analyze the statistics of $H^2<0$ crashes. We use a constant time step in the conformal time domain, $\Delta\tau$, and for three choices of $\alpha$ values calculate the fraction of completed runs (i.e. those that do not encounter an $H^2<0$ crash before the present) for different values of $\Delta \tau$. The results are shown in Figure \ref{fig:crashinganalysis}. Each data point was produced using  10,000 simulations. When $\alpha=0.02$ (not shown in the figure), less than 10 out of 10,000 for any given $\Delta \tau$ do not end up crashing, showing that such values for $\alpha$ are not practical in Model 1 (if we take $S_0=0$).

The results in Figure \ref{fig:crashinganalysis} are for $\alpha = 0.005,\,0.01$ and $0.015$. As anticipated, we see that as $\alpha$ increases, the fraction of completed runs drops considerably.  The trends with $\Delta \tau$ are more visible for the curves corresponding to the two smaller values of $\alpha$, as a large fraction of runs crashed for $\alpha=0.015$. Once again as anticipated from the discussion in the first paragraph of this subsection, as we decrease the size of the time step, the fraction reaching the present day increases (in fact it starts to plateau for the smallest values of $\Delta \tau$).

We also explored a few other types of time steps and observed similar behavior for them. For example, we considered constant time steps in $\tau$, $t$, $a$, $\log a$. etc. In all cases, as we decrease the size of the time steps, the crashing fraction decreases and finally becomes almost constant. 

The $H^2<0$ crashing frequency is also sensitive to the starting time of the simulations. The volume of the past lightcone decreases as we go back in time, and with that the dark energy density can become very large as it is of the order of $1/\sqrt{V}$. For the results we report on, we start the simulations from $z \approx 10^4$, and except in the next subsection, we calculate the volume up to this time using only the matter and the radiation component of the universe. We have seen that if we start increasing the starting redshift, the crashing fraction also increases. In fact, for the analysis shown in Figure \ref{fig1}, where we take a logarithmic time step in $a$ and a starting redshift $z\sim 10^8$, the crashing fraction is almost $100\%$ for $\alpha=0.01$.  

\subsection{Initial Conditions}\label{subsec:ic}

As mentioned at the beginning of Section \ref{subsec:numrec}, we need to choose initial values for the volume, $V_0$, and the action, $S_0$, for Model 1. In this subsection we discuss some of the trends that result from different choices for these initial conditions.

There are several approaches to setting these initial values. Everpresent $\Lambda$ does not tell us about the evolution of the very early universe, before a number-volume correspondence becomes meaningful. A naive approach would be to ignore the limits of validity of the model and to apply it already at the time of the birth of the first few elements. If we do this, and if the initial time step is too small, it turns out that the simulations crash too often, since $\Lambda$ can fluctuate to a negative value too easily and cause $H^2$ to become negative. An alternative approach is to (in a more or less ad hoc manner) attribute a  $V_0$ and $S_0$ to the early non-geometric causal set, when it has presumably reached the size it will begin to grow into an FLRW manifoldlike causal set. From a purely pragmatic point of view, setting $S_0>0$  adds a positive bias to $\Lambda$ that makes it less likely to become too negative, and the simulation less likely to crash. Therefore, $S_0>0$ values are more interesting and we will focus on this case below. Similarly, we will only focus on $V_0>0$ as the spacetime volume and number of elements can only be positive. Note that the initial volume $V_0$ only enters the numerical process described in Section \ref{subsec:numrec} in the denominator of \eqref{lambdai}, by adding it to the calculated volume in the subsequent steps. Therefore, we do not include $V_0$ in the fluctuating numerator part of \eqref{lambdai} which represents the effect of the random action of Model 1. We also omit $V_0$ in our plots of the volume below.

\begin{figure}[H]
\centering
     \includegraphics[width=0.75\textwidth]{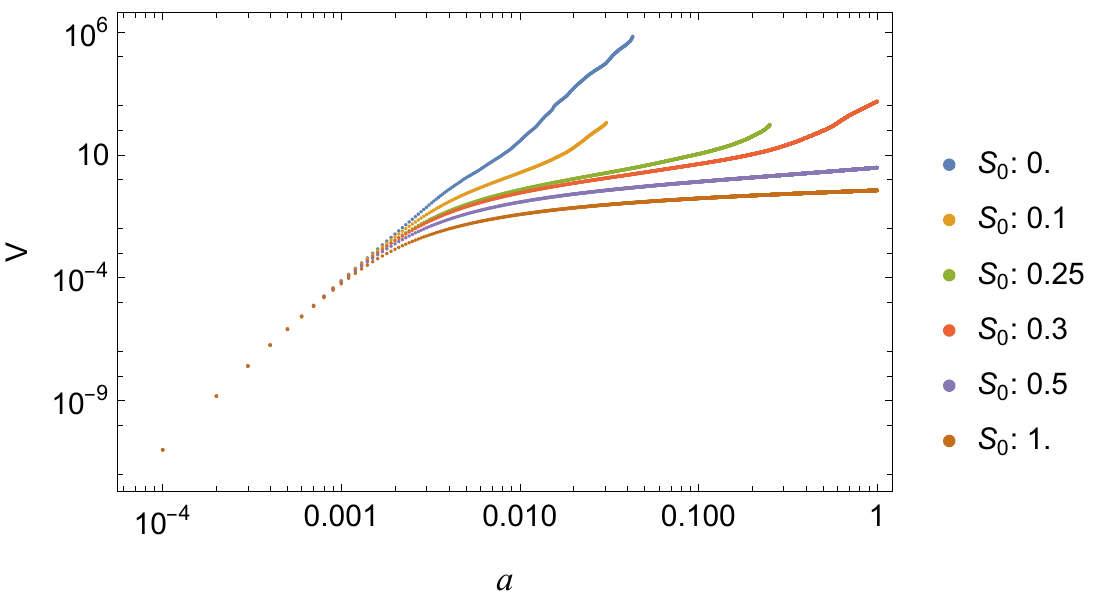}
    \caption{A log-log plot of $V$ (in  arbitrary units) versus $a$ when increasing $S_0$, with $\alpha=0.015$.  The late-time power-law scalings with $a$, in order of increasing $S_0$ are: 6.71, 5.52, 7.82 for the three crashed runs, and 3.37, 0.587, 0.264 for the completed runs.  Compare with the expected scaling of $6$ when $S_0$ is zero and $\alpha$ is small.}
\label{logVlogawithIncreasingS}
\end{figure}

\begin{figure}[H]
\centering
     \includegraphics[width=0.75\textwidth]{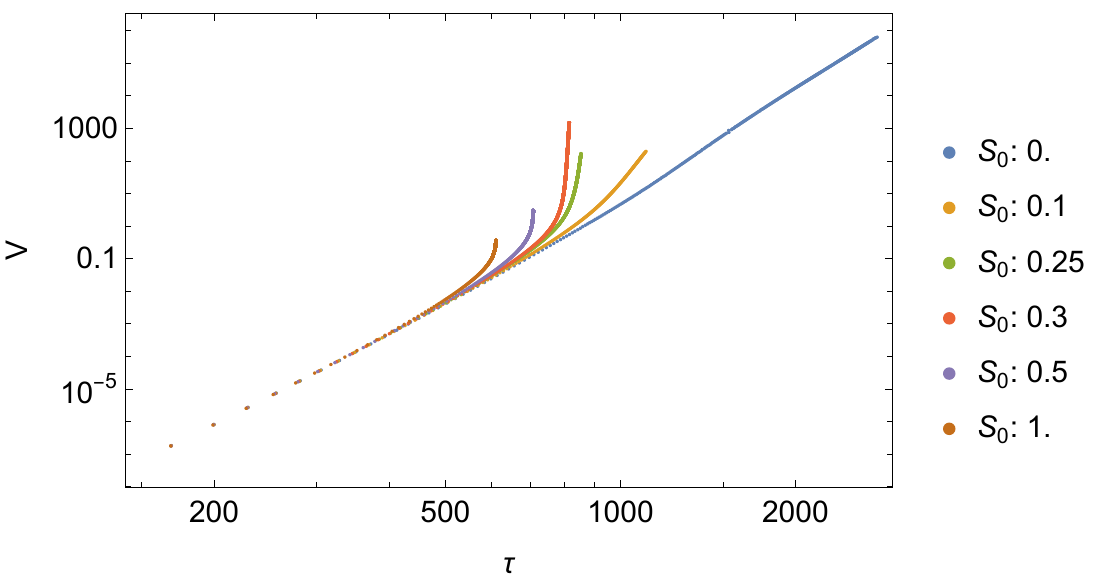}
    \caption{A log-log plot of $V$ versus $\tau$ when increasing $S_0$, with $\alpha=0.015$.  The late-time power-law scalings with $\tau$, in order of increasing $S_0$ are: 11.0, 22.0, 131 for the three crashed runs, and 291, 445, 397 for the completed runs.  Compare with the expected scaling of $12$ when $S_0$ is zero and $\alpha$ is small.  We therefore see that the slow growth of $V$ in units of $a$ in Figure \ref{logVlogawithIncreasingS} is counteracted by the slow growth of $\tau$ with $a$, so that $V$ has an explosive growth in physical time, as one would expect when there is a highly dominant dark energy. Here $V$ is in arbitrary units and $\tau$ is in \text{MPc}.}
\label{logVlogtwithIncreasingS}
\end{figure}

\begin{figure}[H]
\centering
     \includegraphics[width=0.75\textwidth]{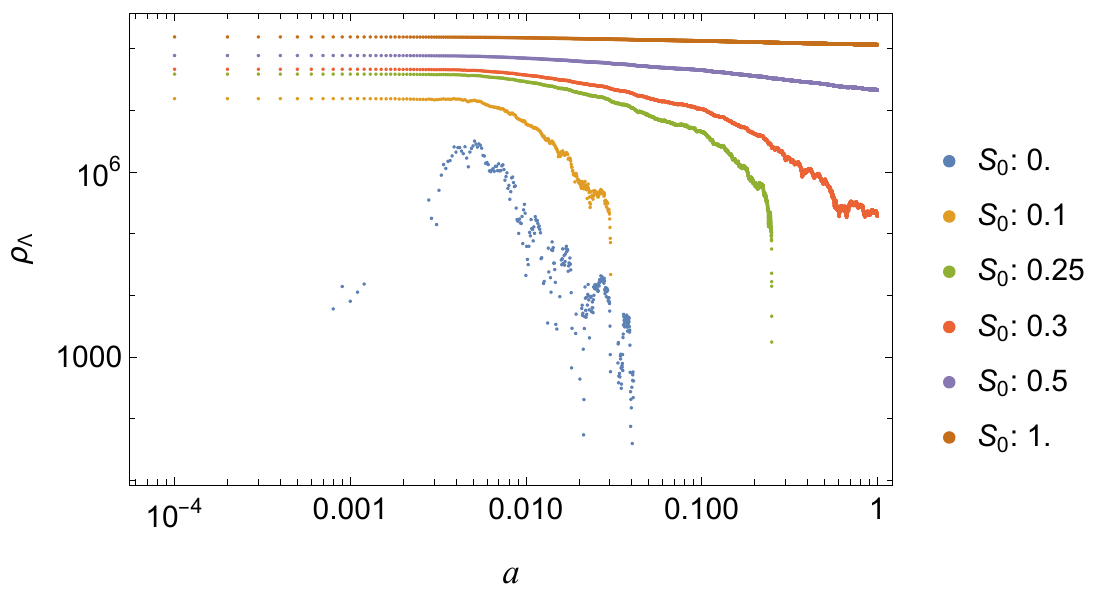}
    \caption{A log-log plot of $\rho_\Lambda$ versus $a$ when increasing $S_0$, with $\alpha=0.015$ .  The expected behaviour of Everpresent $\Lambda$ is that it follows the same scaling as $\rho_m$ at late times, i.e. $\sim a^{-3}$.  For $S_0=0$, $\rho_\Lambda$ does indeed follow that scaling, however the scaling becomes closer to zero, and therefore $\Lambda$ becomes more and more dominant the larger $S_0$ becomes.}
\label{logrholamlogawithIncreasingS}
\end{figure}

\begin{figure}[H]
\centering
     \includegraphics[width=0.8\textwidth]{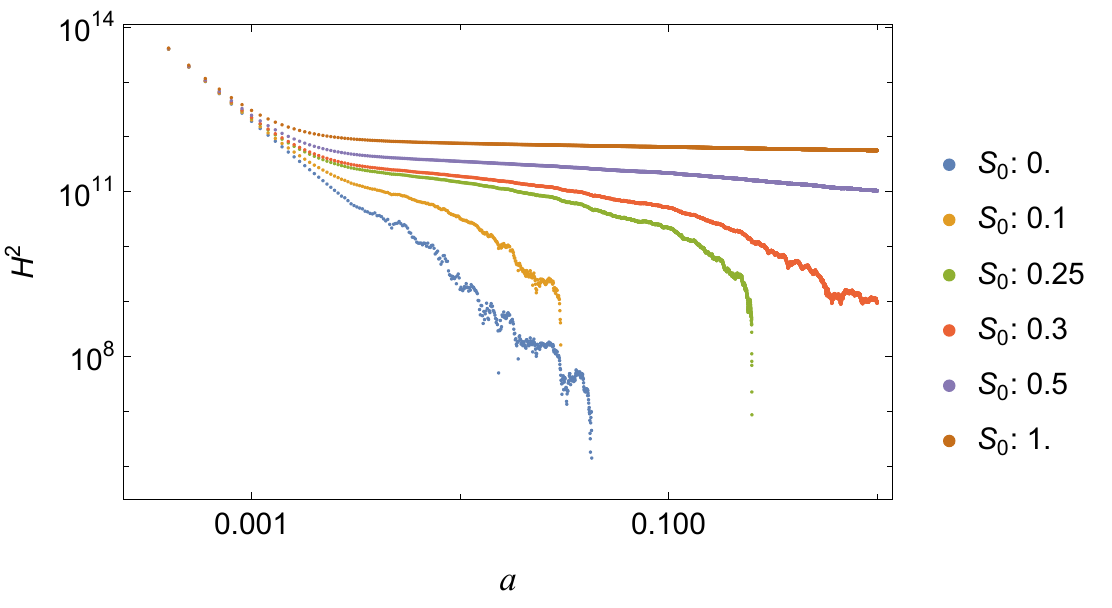}
    \caption{A log-log plot of $H^2$ versus $a$ when increasing $S_0$, with $\alpha=0.015$.  Similar to the case of $\rho_\Lambda$, we expect a decay of $\sim a^{-3}$ from normal Everpresent $\Lambda$ cosmology, which we roughly see for the $S_0$ curve before it crashes.  Furthermore, if the evolution were to predict a value of $H_0$ similar to that of our universe, at $a=1$ we would expect $H^2\sim 10^3$.  However, the completed runs have far higher values, consistent with a large dark energy provided by the initial $S_0$.}
\label{logH2logawithIncreasingS}
\end{figure}

Let us consider the effect of increasing $S_0$. Guided by the results summarised in Figure \ref{fig:crashinganalysis}, we will set $\alpha$ to a value for which most simulations encounter an $H^2<0$  crash, in order to investigate whether a larger $S_0$ can prevent some of these crashes. In Figures \ref{logVlogawithIncreasingS} to \ref{logH2logawithIncreasingS} we show the results. In the simulations we set  $\alpha=0.015$, $V_0=1$, $i_{initial}=1$, $n=10000$, and increase $S_0$ from $0$ to $1$.  We see that as $S_0$ increases, the (initially crashing) simulation eventually stops crashing, although the behaviour  is drastically different in the new regime.  

Although the aim of setting $S_0\neq 0$ was to give a small bias that eventually becomes negligible as $S$ grows, it turns out that the initial value has a persistent effect on the late time behaviour as well. This is because it can take some time for the random fluctuations to have an aggregated effect of the same order
as $S_0$, causing the qualitative fluctuating behaviour in S and $\rho_\Lambda$ to set in at late times and even well after $a=1$.  In Figures \ref{logVlogawithIncreasingS} and \ref{logVlogtwithIncreasingS} we see the scaling of the volume $V$ of the past lightcone with scale factor $a$ and conformal time $\tau$ respectively. We see the curious behaviour that in simulations with smaller $S_0$, while the $H^2<0$  crashing is not prevented, $V$ has approximately the expected scaling with $a$ (a power-law with exponent $6$), whereas simulations with larger $S_0$ have a much smaller scaling, ending at roughly a power of $1/4$.   These late-time scalings are estimated by taking the slope of a straight line fit through the last stretch of the log-log plot.  The larger $S_0$, the more $V$ appears to plateau with $a$, and the smaller the final value of $V$ (for those runs that complete).

While it might seem from Figure \ref{logVlogawithIncreasingS}  that positive $S_0$ has the effect of slowing the expansion of the universe, from Figure \ref{logVlogtwithIncreasingS} we see that the effect is actually the opposite.    We see that while the  scaling of $V$ with $\tau$ at late times for $S_0=0$ is close to the expected power-law with exponent $12$, the larger $S_0$ becomes, the faster the growth of $V$ with $\tau$ becomes.  For example, for $S_0=1$, the scaling of $V$ with $\tau$ is much larger -- close to a power of $400$ -- and the value of $\tau$ is smaller at the end of the simulation.  In other words, the slow growth of $V$ with $a$ is counteracted by the even slower growth of $\tau$ with $a$ (as $\tau$ is also a calculated quantity in Model 1). The reason the final value of $V$ is smaller for larger $S_0$, is that for larger values of $S_0$ the universe becomes more dark energy dominated.  The universe is then in an inflation-like phase with an exponential expansion. Then, $-\tau$ is inversely proportional to $a$ (i.e., the growth of conformal time with $a$ plateaus), and the simulation stops at a smaller $\tau$. Since the final $\tau$ is smaller for larger $S_0$, the final volume of the past lightcone will be smaller, despite its faster growth rate. 
The effect of adding an $S_0>0$ is therefore to add a positive cosmological constant that causes an accelerated expansion, and the effect becomes almost explosive for large enough $S_0$.  This has the further effect of making $\rho_\Lambda$ the dominant contribution to the energy density, as it ceases to follow the late-time $a^{-3}$ scaling of $\rho_m$ observed in Figure \ref{fig1} and ends on a much larger final value.  See Figure \ref{logrholamlogawithIncreasingS}:  The scaling for the $S_0=0$ run has approximately the correct behaviour before crashing, and those runs that do not crash end with much larger final values, as mentioned.  This also has the effect of making $\Omega_\Lambda$ larger, the larger $S_0$ is, and it is effectively 1 (as opposed to the standard value of $\sim 0.7$ for our universe) for the runs that do not encounter an  $H^2<0$  crash.

Finally, the Hubble parameter squared, $H^2$, also increases and ends on unphysical values when $S_0$ is increased.  This is because $H^2$ is proportional to the total energy density $\rho_{tot}$, which becomes dominated by $\rho_\Lambda$ for larger $S_0$.  Simulations that approximate our universe should have final values of $H^2$ that are of order $10^3$, however the values obtained are orders of magnitude larger, as can be seen in Figure \ref{logH2logawithIncreasingS}.

 The results of the simulations above were for a particular value of $\alpha$, and a seed for which the runs reach an $H^2<0$  crash when $S_0=0$.  If we instead consider the behaviour for a smaller $\alpha$ where the simulation with $S_0=0$ does not crash, the effect of increasing $S_0$ leads to similar results to above, except that for intermediate values of $S_0$ the runs will actually encounter an $H^2<0$  crash.  We therefore have 3 ``regimes'' when setting $S_0$: small $S_0$, where the effect on the final results are modest; intermediate $S_0$, where fluctuations become large enough to crash almost all runs\footnote{We can actually see this effect in the figures above, since the example run with $S_0=0.1$ crashes earlier than the run with $S_0=0$.}; and large $S_0$, where $\Lambda$ dominates and we get scalings that are unlike our universe.

It is also interesting to note that $S_0$ and $\alpha$ induce similar behaviour in Model 1. The same observation as above holds for $\alpha$, namely that for small $\alpha$ 
almost all runs succeed, and have the correct scalings.  For intermediate $\alpha$ (larger than $0.015$ when $S_0=0$, as seen in Figure \ref{fig:crashinganalysis}) almost all runs crash.  For large enough $\alpha$, we again obtain similar results to the results for large $S_0$, where fewer simulations crash, and $\rho_\Lambda$ dominates. 

Finally, we have not said anything about the choice of $V_0$.  This is because the simulations do not seem to be sensitive to increasing its value. While we should choose a positive value large enough to keep the runs from initially crashing, the results obtained do not appear to be sensitive to increasing that value further (even by orders of magnitude).

\subsection{Understanding the Auto-Correlation Time} 

An important property of Model 1 is that the auto-correlation time of $\Lambda$ is of the order of the Hubble time. Intuitively, this means that it takes roughly the Hubble time for $\Lambda$ to become independent of its past values and change sign. By one Hubble time, we mean the time it takes for the scale factor to roughly double.\\
To see why this is true, we will calculate the auto-correlation $\langle\Omega_\Lambda(a_1)\Omega_\Lambda(a_2)\rangle$, where  $\Omega_\Lambda=\rho_\Lambda/\rho_{tot}=\Lambda/3H^2$. Combining \eqref{discsum} and \eqref{Slam2} gives
\begin{equation}
    \Lambda(a_1)=\frac{4\pi\ell_p^2}{V(a_1)}\sum_{e_i\in\mathcal{V}(a_1)} b(e_i)\,.
\end{equation}

The summation is performed on all the causal set elements $e_i$ that are in the past lightcone of a point at $a_1$, and $b$ is a Bernoulli random number:
\begin{equation}
    b(e_i)=\begin{cases}
    \phantom{-}1,\ \ \ p=\frac{1}{2}\\
    -1,\ \ \ p=\frac{1}{2}
    \end{cases}.
\end{equation}
We assume the random variables are independent, such that $\langle b(e_i)b(e_j)\rangle=\delta_{ij}$; therefore, if $a_1<a_2$ we have
\begin{equation} \label{lambdacor}
    \langle V_1 \mkern 2mu V_2\,\Lambda(a_1)\mkern 1mu\Lambda(a_2)\rangle=16\pi^2\ell_p^4\sum_{e_i\in\mathcal{V}(a_1)} 1= 16\pi^2\frac{\ell_p^4}{\ell_{cs}^4}\langle V_1\rangle=(8\pi\alpha)^2\langle V_1\rangle,
\end{equation}
where  $V_i\equiv V(a_i)$ and the nonzero contribution to the sum is the common past of the two points which is $\mathcal{V}(a_2)\cap\mathcal{V}(a_1)=\mathcal{V}(a_1)$. 

For obtaining the correlation $\langle\Omega_\Lambda(a_1)\Omega_\Lambda(a_2)\rangle$, we can replace $\Lambda(a) = 3 \Omega_\Lambda H^2$ in \eqref{lambdacor} and assume that we can take $V$ and $H$ in and out of the expectation values. This assumption is certainly not correct, but for the purpose of a rough understanding of the correlation time, it suffices. A few algebraic manipulations lead us to

\begin{equation}
    \langle\Omega_\Lambda(a_1)\,\Omega_\Lambda(a_2)\rangle
    =\frac{(8\pi\alpha/3)^2}{\langle H_1^2\rangle\langle H_2^2\rangle\langle V_2\rangle}\,.
\end{equation}
Next we write $\langle V(a)\rangle=\beta/\langle H(a)^2\rangle^2$, where we expect $\beta\sim\mathcal{O}(10^{-2})$ based on the third column of Figure \ref{fig3:model1_stat}. So we find
\begin{equation}
    \langle\Omega_\Lambda(a_1)\,\Omega_\Lambda(a_2)\rangle=\frac{(8\pi\alpha)^2}{9\beta}\frac{\langle H_2^2\rangle}{\langle H_1^2\rangle}\,.
\end{equation}
Since $\langle\Lambda\rangle=0$, and $\langle H^2\rangle \propto \sum_i \langle\rho_i \rangle$ from the Friedmann equation \eqref{Friedman1},  we find that in the matter dominated era

\begin{equation}
    \langle\Omega_\Lambda(a_1)\,\Omega_\Lambda(a_2)\rangle
     = \frac{(8\pi\alpha)^2}{9\beta}\left(\frac{a_1}{a_2}\right)^3.
\end{equation}
Changing the time variable from the scale factor $a_1$ to $\lambda_1\equiv\ln a_1$ we have shown that
\begin{equation}
\label{autocor}
    \langle\Omega_\Lambda(\lambda_1)\,\Omega_\Lambda(\lambda_2)\rangle=\frac{(8\pi\alpha)^2}{9\beta}e^{-3|\lambda_1-\lambda_2|},
\end{equation}
which suggests that the auto-correlation time is of the order of one Hubble time.

\begin{figure}[H]
\centering
     \includegraphics[width=0.60\textwidth]{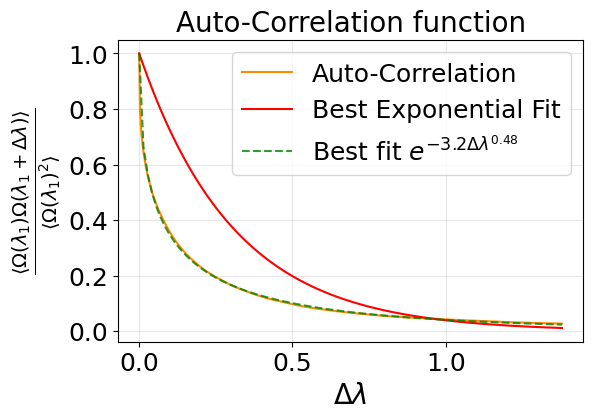}
    \caption{
    The normalized auto-correlation function $\frac{\langle\Omega_\Lambda(\lambda_1)\,\Omega_\Lambda(\lambda_1+\Delta\lambda)\rangle}{\langle\Omega_\Lambda(\lambda_1)^2\rangle}$  versus $\Delta \lambda$.  To make this plot, 
    3000 realizations of Model 1 were simulated with the parameters $\alpha=0.01$ and  $\Omega_m^0 h^2=0.14$. For each simulation, $\langle\Omega_\Lambda(\lambda_1)\,\Omega_\Lambda(\lambda_1+\Delta\lambda)\rangle$ was calculated by averaging over $\lambda_1$. The resulting function of $\Delta\lambda$ was then averaged over the 3000 realizations, and suitably normalized so that it equals 1 at $\Delta\lambda=0$. The final auto-correlation function is the orange curve, and the red curve is the best fit of an exponential $e^{-c|\lambda_1-\lambda_2|}$ to it. The dashed curve is the best fit of a function of the form $e^{-c_1|\lambda_1-\lambda_2|^{c_2}}$.}
\label{autocorfig}
\end{figure}

We have made some naive assumptions in deriving \eqref{autocor}, hence we do not expect the auto-correlation function to be exactly an exponential function. In Figure \ref{autocorfig} we have plotted the auto-correlation of $\Omega_\Lambda$ (for a given $\alpha$) versus $\Delta \lambda$ along with a best-fit exponential function.  Indeed, one can see that the auto-correlation function is not an exponential. The best fit to the orange curve turns out to be approximately $e^{-3.2 (\Delta \lambda)^{0.48}}$.

\begin{figure}[H]
    \centering
    \begin{subfigure}[h]{0.6\textwidth}
         \centering
         \includegraphics[width=\textwidth]{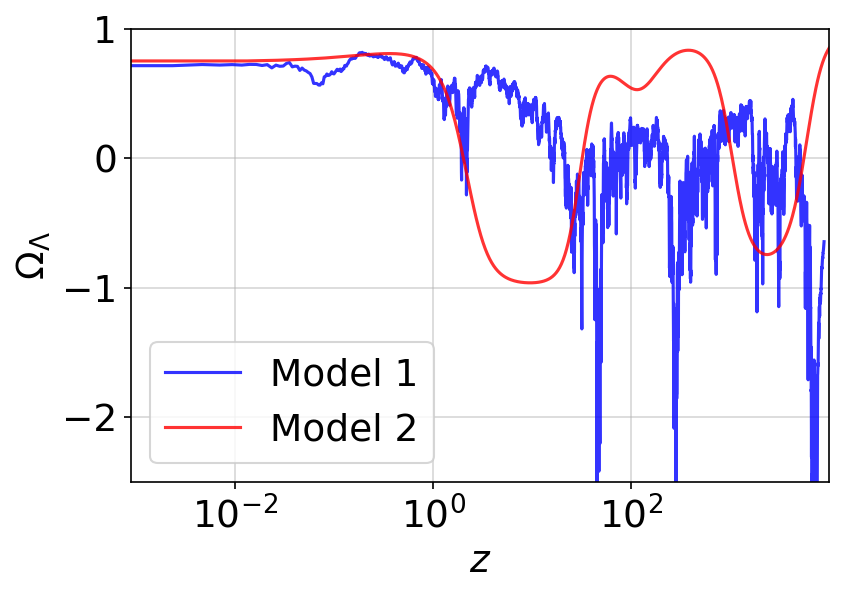}
     \end{subfigure}
 \caption{Plot of $\Omega_\Lambda$ versus redshift for one realization of Model 1 (in blue) and one realization of Model 2 (in red). The seeds for this figure were chosen such that for the given set of parameters of the models ($\alpha=0.01$, $\Omega_m^0h^2=0.14$ for Model 1 and $\Tilde{\alpha}=1.0$, $\mu=1.108$ for Model 2), $\Omega_\Lambda^0\sim0.7$; hence they are not necessarily among the typical realizations of the models. 
 }
\label{fig4:m1m2}
\end{figure}

Nevertheless, the dependence on $\Delta\lambda$ observed in the simulation data confirms that when the scale factor is multiplied by an $\mathcal{O}(1)$ number, the correlation of $\Lambda$ with its previous values gets highly suppressed. Also note that the auto-correlation curve of Model 1 is dropping very fast close to $\Delta\lambda\sim0$. This means that in very small time scales, $\Omega_\Lambda$ fluctuates a lot with a small magnitude. This is why the plot of $\Omega_\Lambda$ (Figure \ref{fig4:m1m2}, in blue) looks jagged at small intervals.

\subsubsection{Model 2}
An alternative phenomenological model of Everpresent $\Lambda$, known as Model 2, was put forth in \cite{Zwane_2018}. Model 2 is designed to incorporate the fluctuating nature of dark energy in Everpresent $\Lambda$ as well as the Hubble time auto-correlation emergent in Model 1. This is done by making the assumption that $-1\leq\Omega_\Lambda(\lambda)\leq1$,
which allows for the definition of a new variable  $\hat{\Omega}_\Lambda(\lambda)=\tanh^{-1}(\Omega_\Lambda(\lambda))$. In fact, $\Omega_\Lambda(\lambda)\leq1$ is always true since the other components of the universe have positive energy density, but $\Omega_\Lambda(\lambda)\geq-1$ is an assumption that is made to allow for the convenient definition of an $\hat{\Omega}_\Lambda$ that can be generated by a Gaussian process.
Given that $\hat{\Omega}_\Lambda(\lambda)$ is unbounded, it can be generated as a \emph{Gaussian process}.  Over a discrete domain, this is done by treating the set of $\hat{\Omega}_\Lambda$ at different $\lambda_i$'s as variables over which a multivariate Gaussian distribution is defined.  The random vector drawn from this distribution then describes the variation of $\hat{\Omega}_\Lambda$ over $\lambda$.  The precise way in which  $\hat{\Omega}_\Lambda$ auto-correlates is dictated by the covariance matrix $\langle\hat{\Omega}_\Lambda(\lambda_1)\hat{\Omega}_\Lambda(\lambda_2)\rangle$ in the Gaussian distribution. Hence, one can instantly generate a history of $\Omega_\Lambda(\lambda)$ using the covariance matrix of $\hat{\Omega}_\Lambda(\lambda)$. As this model does not involve volume calculations, it is computationally much faster than Model 1. In \cite{Zwane_2018} the ansatz was made that the covariance matrix is a squared-exponential
\begin{equation}
\label{model2}   \langle\hat{\Omega}_\Lambda(\lambda_1)\hat{\Omega}_\Lambda(\lambda_2)\rangle=\Tilde{\alpha}^2e^{-\frac{\mu}{2}(\lambda_1-\lambda_2)^2},
\end{equation}
where $\Tilde{\alpha}$ and $\mu$ are free parameters of the model. 

 For $\mu\sim\mathcal{O}(1)$, the auto-correlation time of the fluctuations in dark energy are of the order of one Hubble time.\footnote{Even though the covariance matrix in Model 2 is for $\hat{\Omega}_\Lambda$, for values of $\tilde{\alpha}\approx 1$, the auto-correlation of $\Omega_\Lambda(\lambda)$ closely follows that of $\hat{\Omega}_\Lambda(\lambda)$.} However, one essential difference between the realizations of Model 1 and those of Model 2 is that the histories of $\Omega_\Lambda$ in Model 1 are more jagged at smaller time scales (see Figure  \ref{fig4:m1m2}). The reason is that the square root exponential auto-correlation function of Model 1 drops faster close to $\Delta\lambda=0$.\\
Taking these considerations into account, more models can be built based on the Everpresent $\Lambda$ idea that are computationally as fast as Model 2 but better resemble the realizations of Model 1. For this purpose, \eqref{model2} should be replaced with a square root exponential function, or even with the numerical orange curve from Figure \ref{autocorfig}.

\section{Discussion\label{sec:discussion}}

We have laid out and reviewed the foundations of Everpresent $\Lambda$ cosmology. We also investigated  key properties of simulations based on the original phenomenological implementation of it which is in terms of a random action per causal set element. 

 We found that in Model 1 the value of $H_0$ for the noncrashing simulations is approximately normally distributed about a mean that is close to the value it would have if $\Lambda$ were zero.  While for small values of the fluctuation parameter $\alpha$ the observed value of $H_0$ is highly unlikely, we found that increasing this parameter increases the width of the distribution of $H_0$, and today's value becomes more likely. 
 
 We further found that Model 1 appears to have three ``phases''.  A phase with small $\alpha$ and $S_0$ where the behaviour is close to the standard behaviour with a small cosmological constant, one with intermediate $\alpha$ and/or $S_0$ where the fluctuations are too large and most runs encounter an $H^2<0$ crash, and a new phase with large $\alpha$ and/or $S_0$, where many seeds that crashed in the intermediate phase cease to crash.  In this third phase we see rapid growth for the spacetime volume, and a dark energy density that no longer tracks the matter density. An interesting avenue for future work is to investigate whether by tuning the parameters $S_0$ and $\alpha$, we could have an inflation-like phase in the early universe with a graceful exit to radiation domination. Some signs of a slow-roll type of behaviour and then an exit to a decreasing Hubble parameter can already be seen in Figure~\ref{logH2logawithIncreasingS}.
 
 Finally, we found that in Model 1 the sign of $\Lambda$ is likely to change roughly every Hubble time.  An observational evidence of sign change in dark energy would therefore be a strong indication of Everpresent $\Lambda$ in our universe.

While we have aimed to provide a complete exposition of the topics relevant to Everpresent $\Lambda$, there are a number of things we did not delve into. Let us briefly mention three of these: (1) One of the  assumptions of the model is that the mean about which $\Lambda$ fluctuates is $0$; motivating this assumption with a Lorentzian argument remains an open question. Ideas from the Lorentzian path integral~\cite{Feldbrugge:2017kzv} and analogue gravity~\cite{Samuel:2006ga} can potentially be fruitful in this direction. (2) There are several classical aspects of the model that likely must be replaced by their quantum counterparts. Of course this is easier said than done, in the absence of the complete quantum gravity theory. However, there are hints at how this can begin to be explored. For example, one can replace the classical stochastic random walk of Model 1 with a quantum random walk inspired by \cite{Martin:2004xi, Venegas-Andraca:2012zkr}. See also \cite{Will1, Will2}. As well, insights from quantum unimodular gravity \cite{Padilla:2014yea, Eichhorn:2013xr} can be consulted for indications of a modified dynamics. (3) Mathematically, it may be worth doing an analysis of the Friedmann equations, treating them as stochastic differential or difference equations, in which $\Lambda$ is the random component.  Such an analysis may offer insights on the statistical evolution of the scale factor as well as how to handle negative energy densities and derivatives of $\Lambda$.

It would also be interesting to explore connections to \cite{Wang:2017oiy, Cree:2018mcx} where a stochastic cosmological constant from fluctuations of vacuum energy was studied.

In another paper we will report on our results from using Models 1 and 2 to confront supernova and cosmic microwave background data. It is perhaps a bit premature to test these simple models against the full range of data, but we can nevertheless learn from the analysis. The details of this work can be found in \cite{das2023aspects}.

As mentioned in the introduction, many of our conventional theoretical models have fallen short of shedding light on cosmological open questions. The cosmological constant puzzle has been a longstanding example of such an open question, and more recently, there have emerged new puzzles such as the Hubble tension. Moreover, in an era where \emph{understanding} cosmological data (which will, in turn, relieve tensions) takes priority over merely fitting it, we believe that models such as Everpresent $\Lambda$ should be taken seriously and further developed and explored.

\section*{Acknowledgments}We thank Niayesh Afshordi, Fay Dowker, and Rafael Sorkin for several helpful discussions and suggestions. We also thank Steve Carlip for helpful comments on the manuscript. This work was partially funded by a Leverhulme Trust Research Project Grant. YY acknowledges financial support from Imperial College London through an Imperial College Research Fellowship grant. AN is funded by the President’s PhD Scholarships from Imperial College London.

\appendix
\section{On the Equations of Motion of Global Unimodular Gravity}\label{appendixA}
\subsection{An Alternative Derivation of the Equations of Motion in Unimodular Gravity}

In order to find the stationary point of the action in \eqref{Zglobal1}, let us perform the variation $g_{\mu\nu}\rightarrow g_{\mu\nu}+\delta g_{\mu\nu}$. The volume constraint in the path integral gives
\begin{equation}
\label{constraintG}
    \int d^4x \sqrt{-g}=V\ \ \Rightarrow\ \ \int d^4x\sqrt{-g}\ \delta g_{\mu\nu}g^{\mu\nu}=0.
\end{equation}

Now the stationary-action principle requires $\delta S=0$. On the other hand, we know
\begin{equation}
\label{variedS}
    \delta S\propto \int d^4x\sqrt{-g}\ \delta g_{\mu\nu}E^{\mu\nu},
\end{equation}
where $E_{\mu\nu}=G_{\mu\nu}-8\pi G\mkern 2mu T_{\mu\nu}+\tilde{\Lambda} g_{\mu\nu}$. 
Note that we can no longer infer the full Einstein equations $E_{\mu\nu}=0$ from \eqref{variedS} because $\delta g_{\mu\nu}$ should also satisfy \eqref{constraintG}. The condition  \eqref{constraintG} is satisfied if we rewrite the metric $g_{\mu\nu}$ in terms of an unconstrained metric $\hat{g}_{\mu\nu}$ as follows\footnote{In the variational approach one normally obtains an equation of motion by setting the coefficient of the varied function to zero.  Here, since we have an additional constraint, it turns out to be helpful to define our function $g$ in terms of an auxiliary function $\hat{g}$ such that $g$ satisfies the constraint, while the coefficient of $\hat{g}$ in the varied action gives the equation of motion. Notice that since $\hat{g}$ is defined by a rescaling of $g$, we will have relations such as $\hat{g}^{\alpha\beta}(y)\hat{g}_{\mu\nu}(x)=g^{\alpha\beta}(y)g_{\mu\nu}(x)$.}
\begin{equation}
\label{gghat}
    g_{\mu\nu}(x)=\sqrt{\frac{V}{\int d^4y \sqrt{-\hat{g}(y)}}}\ \hat{g}_{\mu\nu}(x).
\end{equation}

 Performing a variation on \eqref{gghat} gives
\begin{equation} \label{gvar}
    \delta g_{\mu\nu}(x)=\sqrt{\frac{V}{\int d^4 y\sqrt{-\hat{g}(y)}}}\left(\delta\hat{g}_{\mu\nu}(x)-\frac{1}{4}\frac{\int d^4y\sqrt{-\hat{g}(y)}\ \delta\hat{g}_{\alpha\beta}(y)g^{\alpha\beta}(y)}{\int d^4 y\sqrt{-\hat{g}(y)}}g_{\mu\nu}(x)\right).
\end{equation}

Then for finding the classical equations of motion, we substitute \eqref{gvar} into \eqref{variedS} and then set the coefficient of $\delta\hat{g}_{\mu\nu}(x)$ in $\delta S$ equal to zero. A bit of algebra reveals
\begin{equation}
\label{EEE}
    E_{\mu\nu}(x)-\frac{1}{4}g_{\mu\nu}(x)\langle E\rangle=0,
\end{equation}
where $\langle E\rangle$ is the spacetime average of the trace of $E_{\mu\nu}$:
\begin{equation}
    \langle E\rangle=\frac{\int d^4y\sqrt{-g}\ E(y)}{\int d^4y\sqrt{-g}}.
\end{equation}
For an alternative derivation of \eqref{EEE}, see Appendix \ref{EEEderivation}. Taking the trace of \eqref{EEE} reveals that $E=\langle E\rangle$, i.e., $E$ must be a constant:
\begin{equation}
\label{cons}
    -R-8\pi G\mkern 2mu T+4\tilde{\Lambda}= \text{const}.
\end{equation}
In particular, \eqref{EEE} can now be rewritten as
\begin{equation}
\label{EE}
    E_{\mu\nu}(x)-\frac{1}{4}g_{\mu\nu}(x)E=0\ \ \Rightarrow \ \ G_{\mu\nu}-8\pi G \mkern 2mu T_{\mu\nu}+\frac{1}{4}g_{\mu\nu}(R+8\pi G\mkern 2mu T)=0,
\end{equation}
 
and as argued in Section \ref{subsec:ug}, $(R+8\pi G\mkern 2mu T)$ must be constant, and can be set to $4\Lambda$.  We therefore obtain 
\begin{equation}
    G_{\mu\nu}+\Lambda g_{\mu\nu}=8\pi G\mkern 2mu T_{\mu\nu}\,,
\end{equation}
as we did in \eqref{lugeom}.

\subsection{An Alternative Derivation of \eqref{EEE}} \label{EEEderivation}
In the usual variational principle, say for the Einstein-Hilbert action, one can argue that since $\delta g_{\mu\nu}$ is arbitrary, it can be chosen to be $\epsilon\delta^{\mu_0}_{\mu}\delta^{\nu_0}_{\nu}\delta(x-x_0)$, where $\epsilon$ is some small number. Then from \eqref{variedS}, the stationary condition $\delta S\propto \int d^4x\sqrt{-g}\ \delta g_{\mu\nu}E^{\mu\nu}=0$ gives $E^{\mu_0\nu_0}(x_0)=0$, which is the Einstein equation for any point $x_0$ and for any component $\mu_0\nu_0$. In unimodular gravity, however, $\delta g_{\mu\nu}$ is not totally arbitrary. 

One might naively think that we can satisfy \eqref{constraintG}, i.e.,
\begin{equation}
\label{constraintGappendix}
    \int d^4x\sqrt{-g}\ \delta g_{\mu\nu}g^{\mu\nu}=0,
\end{equation}
by making a change in $\delta g_{\mu\nu}$ at infinity to compensate for a delta function at the point $x_0$, and then conclude that the same equation of motion as GR can be derived from global unimodular gravity. However, if the variation at infinity is done with care, it will lead to the same conclusion as equation \eqref{EEE} (i.e., $E_{\mu\nu}(x)-\frac{1}{4}g_{\mu\nu}(x)\langle E\rangle=0$). Let us expand on this point.
To satisfy \eqref{constraintG}, we can choose a point $x_1$ far away from $x_0$ and set
\begin{equation}
    \delta g_{\mu\nu}=\epsilon\left(\delta^{\mu_0}_{\mu}\delta^{\nu_0}_{\nu}\delta(x-x_0) - \delta^{\mu_1}_{\mu}\delta^{\nu_1}_{\nu}\delta(x-x_1)\frac{g^{\mu_0\nu_0}(x_0)}{g^{\mu_1\nu_1}(x_1)}\right).
\end{equation}
It is easy to check that this variation satisfies \eqref{constraintG}. Now $\delta S$ should be zero for such a variation. This gives
\begin{equation}
    E^{\mu_0\nu_0}(x_0)-E^{\mu_1\nu_1}(x_1)\frac{g^{\mu_0\nu_0}(x_0)}{g^{\mu_1\nu_1}(x_1)}=0.
\end{equation}
This may seem to be a nonlocal equation, since it is relating tensors that can be spacelike separated. But in fact it is just telling us that $\frac{E^{\mu_0\nu_0}(x_0)}{g^{\mu_0\nu_0}(x_0)}$ is a constant which does not depend on $x_0$ or $\mu_0\nu_0$. This is exactly the statement of \eqref{EEE}. It is equivalent to the statement $E_{\mu\nu}-\frac{1}{4}g_{\mu\nu}E=0$, which is the traceless Einstein equation. The argument that starts from \eqref{gghat} is essentially a more formal derivation of this fact.

\section{On the Fluid Interpretation}\label{subsec:fluid}

In our implementation of Model 1 to create background histories for Everpresent $\Lambda$, we only used the first Friedmann equation \eqref{Friedman1}, which contains the energy density $\rho_\Lambda$ but not the pressure $p_\Lambda$. However, for doing perturbation theory (which we discuss in \cite{das2023aspects}), it is necessary to work with 
the full Einstein equations, which contain both $\rho_\Lambda$ and $p_\Lambda$. As mentioned in Section \ref{subsec:frw}, if we insist on Everpresent $\Lambda$ being indeed a “cosmological constant" in the sense that $p_\Lambda=-\rho_\Lambda$, then the fluctuating $\Lambda$ makes the Friedmann equations \eqref{Friedman1} and \eqref{Friedman2}, and the continuity equation $\Dot{\rho}_\Lambda+3H(\rho_\Lambda+p_\Lambda)=0$ inconsistent with each other. As first suggested in \cite{Ahmed_2013}, one way out of this dilemma  is to relax the condition $p_\Lambda=-\rho_\Lambda$, and allow the equation of state parameter $w_\Lambda\equiv p_\Lambda/\rho_\Lambda$ to be realization-dependent. By this we mean that the time-dependences of $p_\Lambda$ and $\rho_\Lambda$ will be different in each realization of Everpresent $\Lambda$, and will therefore yield a different time-dependence for $w_\Lambda$. This is the fluid interpretation of Everpresent $\Lambda$.

There are two options we have for modifying  the equation of state in order to satisfy all three equations: 

\begin{itemize}
    \item Scheme 1: Set $\rho_\Lambda=\frac{1}{8\pi G}\Lambda$. The continuity equation then implies that \begin{equation}
        \label{sch1}
        p_\Lambda=-\frac{1}{8\pi G}\left(\frac{\Dot{\Lambda}}{3H}+\Lambda\right).
    \end{equation}
    \item Scheme 2: Set $p_\Lambda=-\frac{1}{8\pi G}\Lambda$. By the continuity equation, $\rho_\Lambda$ should then be the solution to the first order differential equation
    \begin{equation}
    \label{sch2}
        \frac{\Dot{\rho}_\Lambda}{3H}+\rho_\Lambda=\frac{\Lambda}{8\pi G}\, .
    \end{equation}
\end{itemize}
In both \eqref{sch1} and \eqref{sch2}, the derivative is with respect to the comoving time $t$.

An advantage of the fluid interpretation is that there is no need to alter the Einstein equations, and hence it may seem that the perturbation theory for the Everpresent $\Lambda$ cosmology would be more straightforward. However, we shall now see that both of the above schemes are problematic.

\begin{figure}[H]
\centering
     \includegraphics[width=0.60\textwidth]{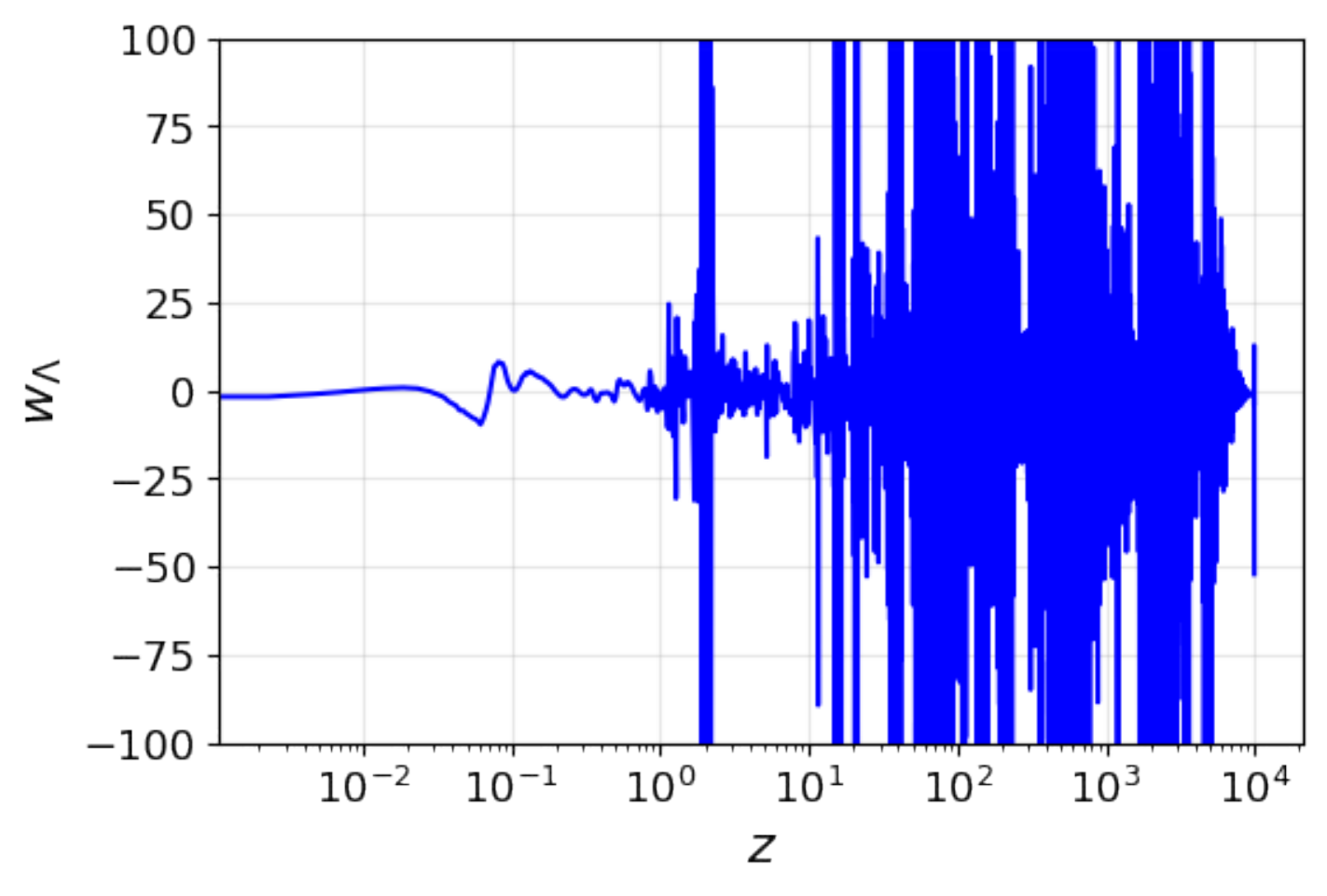}
    \caption{We show  $w_\Lambda$ (in Scheme 1) versus redshift for one realization of Everpresent $\Lambda$ with parameters $\alpha=0.01$, $\Omega_mh^2=0.14$. Each divergence of $w_\Lambda$ corresponds to a zero-crossing of $\Lambda$. (Note that the value obtained for $w_\Lambda$ is dependent on the method used to differentiate the stochastic time series of $\Lambda$).}
\label{wfig}
\end{figure}

In Scheme 1, whenever $\Lambda$ crosses zero, we have $\rho_\Lambda=0$. However, from \eqref{sch1}, $\Dot{\Lambda}$ being non-zero implies that $p_\Lambda\neq0$. Therefore, the equation of state parameter $w_\Lambda$ diverges at these zero-crossing instants, as can be seen in Figure \ref{wfig}. This is not necessarily an issue from the point of view of the background evolutions, but problems arise at the level of perturbations. For example, the equation governing the density contrast of the dark energy fluid, $\delta_\Lambda$, is

\begin{equation}\label{densitycontrast}
\begin{aligned}[b]
    \delta_\Lambda ' +(1+w_\Lambda)(\theta+\frac{1}{2}h')+3\mathcal{H}\left(\frac{\delta p_\Lambda}{\delta\rho_\Lambda}-w_\Lambda\right)\delta_\Lambda=0
\end{aligned}
\end{equation}
The above equation is written in the synchronous gauge. $\delta_\Lambda\equiv\frac{\delta\rho_\Lambda}{\rho_\Lambda}$ is the density contrast of  $\Lambda$ ($\delta\rho_\Lambda$ being the density perturbation), and $\delta p _\Lambda$ is the pressure perturbation of the fluid Everpresent $\Lambda$. Now, according to \eqref{densitycontrast}, a divergence in $w_\Lambda$ is detrimental to the perturbation variable $\delta_\Lambda$. Whenever $\Lambda$ crosses zero, the perturbation theory of Scheme 1 breaks down.

Scheme 2 is problematic in a different way. As shown in \cite{Ahmed_2013}, solving the first Friedmann equation \eqref{Friedman1} with $\rho_\Lambda$ from \eqref{sch2} is essentially equivalent to solving the second Friedmann equation \eqref{Friedman2}. The issue with this is numerical instability. Small initial numerical errors grow uncontrollably and render the numerical integration unreliable. In addition, the problem of divergences in $w_\Lambda$ persists in this scheme at zero-crossings of $\rho_\Lambda$ instead of $\Lambda$.

In our papers, we do not take the fluid interpretation. We try to remain agnostic by only employing the first Friedmann equation for the background, and not adding any perturbation equations for Everpresent $\Lambda$. A completely consistent treatment (that is potentially derived from modifications to the Einstein equations to accommodate a stochastic $\Lambda$) remains to be investigated in the future.

\bibliographystyle{jhep}
\bibliography{EverLam1.bib}

\providecommand{\href}[2]{#2}\begingroup\raggedright\begin{thebibliography}{10}

\bibitem{Cohen:1998zx}
A.~G. Cohen, D.~B. Kaplan, and A.~E. Nelson, {\it {Effective field theory,
  black holes, and the cosmological constant}},  {\em Phys. Rev. Lett.} {\bf
  82} (1999) 4971--4974, [\href{http://arxiv.org/abs/hep-th/9803132}{{\tt
  hep-th/9803132}}].

\bibitem{Freidel:2022ryr}
L.~Freidel, J.~Kowalski-Glikman, R.~G. Leigh, and D.~Minic, {\it {The Vacuum
  Energy Density and Gravitational Entropy}},
  \href{http://arxiv.org/abs/2212.00901}{{\tt arXiv:2212.00901}}.

\bibitem{Bramante:2019uub}
J.~Bramante and E.~Gould, {\it {Material matter effects in gravitational UV/IR
  mixing}},  {\em Phys. Rev. D} {\bf 101} (2020), no.~8 084022,
  [\href{http://arxiv.org/abs/1910.07905}{{\tt arXiv:1910.07905}}].

\bibitem{Krotov:2010ma}
D.~Krotov and A.~M. Polyakov, {\it {Infrared Sensitivity of Unstable Vacua}},
  {\em Nucl. Phys. B} {\bf 849} (2011) 410--432,
  [\href{http://arxiv.org/abs/1012.2107}{{\tt arXiv:1012.2107}}].

\bibitem{Afshordi:2022ive}
N.~Afshordi and J.~a. Magueijo, {\it {Lower bound on the cosmological constant
  from the classicality of the early Universe}},  {\em Phys. Rev. D} {\bf 106}
  (2022), no.~12 123518, [\href{http://arxiv.org/abs/2209.07914}{{\tt
  arXiv:2209.07914}}].

\bibitem{Perez:2018wlo}
A.~Perez, D.~Sudarsky, and J.~D. Bjorken, {\it {A microscopic model for an
  emergent cosmological constant}},  {\em Int. J. Mod. Phys. D} {\bf 27}
  (2018), no.~14 1846002, [\href{http://arxiv.org/abs/1804.07162}{{\tt
  arXiv:1804.07162}}].

\bibitem{Bernard:2012nv}
D.~Bernard and A.~LeClair, {\it {Scrutinizing the Cosmological Constant Problem
  and a possible resolution}},  {\em Phys. Rev. D} {\bf 87} (2013), no.~6
  063010, [\href{http://arxiv.org/abs/1211.4848}{{\tt arXiv:1211.4848}}].

\bibitem{originallambda}
R.~D. Sorkin, {\it {A Modified Sum-Over-Histories for Gravity reported in the
  article by D. Brill and L. Smolin: “Workshop on quantum gravity and new
  directions”}},  in {\em Highlights in gravitation and cosmology:
  Proceedings of the International Conference on Gravitation and Cosmology,
  Goa, India, 14–19 December 1987} (B.~R. Iyer, A.~Kembhavi, J.~V. Narlikar,
  and C.~V. Vishveshwara, eds.), pp.~184--186, 1988.

\bibitem{Sorkin:2007bd}
R.~D. Sorkin, {\it {Is the cosmological 'constant' a nonlocal quantum residue
  of discreteness of the causal set type?}},  {\em AIP Conf. Proc.} {\bf 957}
  (2007), no.~1 142--153, [\href{http://arxiv.org/abs/0710.1675}{{\tt
  arXiv:0710.1675}}].

\bibitem{sorkin1994role}
R.~D. Sorkin, {\it Role of time in the sum-over-histories framework for
  gravity},  {\em International journal of theoretical physics} {\bf 33}
  (1994), no.~3 523--534.

\bibitem{SupernovaSearchTeam:1998fmf}
{\bf Supernova Search Team} Collaboration, A.~G. Riess et~al., {\it
  {Observational evidence from supernovae for an accelerating universe and a
  cosmological constant}},  {\em Astron. J.} {\bf 116} (1998) 1009--1038,
  [\href{http://arxiv.org/abs/astro-ph/9805201}{{\tt astro-ph/9805201}}].

\bibitem{SupernovaCosmologyProject:1998vns}
{\bf Supernova Cosmology Project} Collaboration, S.~Perlmutter et~al., {\it
  {Measurements of $\Omega$ and $\Lambda$ from 42 high redshift supernovae}},
  {\em Astrophys. J.} {\bf 517} (1999) 565--586,
  [\href{http://arxiv.org/abs/astro-ph/9812133}{{\tt astro-ph/9812133}}].

\bibitem{ahmed2004everpresent}
M.~Ahmed, S.~Dodelson, P.~B. Greene, and R.~Sorkin, {\it {Everpresent
  $\Lambda$}},  {\em Physical Review D} {\bf 69} (2004), no.~10 103523.

\bibitem{Ahmed_2013}
M.~Ahmed and R.~D. Sorkin, {\it {Everpresent $\Lambda$ {II}: Structural
  stability}},  {\em Physical Review D} {\bf 87} (mar, 2013).

\bibitem{Zwane_2018}
N.~Zwane, N.~Afshordi, and R.~D. Sorkin, {\it {Cosmological tests of
  Everpresent $\Lambda$}},  {\em Classical and Quantum Gravity} {\bf 35} (sep,
  2018) 194002, [\href{http://arxiv.org/abs/1703.06265}{{\tt
  arXiv:1703.06265}}].

\bibitem{Barrow:2006vy}
J.~D. Barrow, {\it {A Strong Constraint on Ever-Present Lambda}},  {\em Phys.
  Rev. D} {\bf 75} (2007) 067301,
  [\href{http://arxiv.org/abs/gr-qc/0612128}{{\tt gr-qc/0612128}}].

\bibitem{Zuntz:2007sve}
J.~A. Zuntz, {\it {The cosmic microwave background in a causal set universe}},
  {\em Phys. Rev. D} {\bf 77} (2008) 043002,
  [\href{http://arxiv.org/abs/0711.2904}{{\tt arXiv:0711.2904}}].

\bibitem{rafael_talk2}
R.~Sorkin, {\it Supplementary considerations on everpresent lambda},  nov,
  2019.
\newblock PIRSA:19110069 see, \url{https://pirsa.org/19110069}.

\bibitem{bombelli1987space}
L.~Bombelli, J.~Lee, D.~Meyer, and R.~D. Sorkin, {\it Space-time as a causal
  set},  {\em Physical review letters} {\bf 59} (1987), no.~5 521.

\bibitem{tHooft1979}
G.~'t~Hooft, {\em Quantum Gravity: A Fundamental Problem and Some Radical
  Ideas}, pp.~323--345.
\newblock Springer US, Boston, MA, 1979.

\bibitem{Myrheim:293594}
J.~Myrheim, {\it {Statistical geometry}},  tech. rep., CERN, Geneva, 1978.

\bibitem{Sorkin:2003bx}
R.~D. Sorkin, {\it {Causal sets: Discrete gravity}},  in {\em {School on
  Quantum Gravity}}, pp.~305--327, 9, 2003.
\newblock \href{http://arxiv.org/abs/gr-qc/0309009}{{\tt gr-qc/0309009}}.

\bibitem{Dowker:2005tz}
F.~Dowker, {\em {Causal sets and the deep structure of spacetime}},
  pp.~445--464.
\newblock 2005.
\newblock \href{http://arxiv.org/abs/gr-qc/0508109}{{\tt gr-qc/0508109}}.

\bibitem{Henson:2006kf}
J.~Henson, {\it {The Causal set approach to quantum gravity}},
  \href{http://arxiv.org/abs/gr-qc/0601121}{{\tt gr-qc/0601121}}.

\bibitem{surya2019causal}
S.~Surya, {\it The causal set approach to quantum gravity},  {\em Living
  Reviews in Relativity} {\bf 22} (2019), no.~1 1--75.

\bibitem{Henson:2016piq}
J.~Henson, D.~Rideout, R.~D. Sorkin, and S.~Surya, {\it {Onset of the
  Asymptotic Regime for (Uniformly Random) Finite Orders}},  {\em Exper. Math.}
  {\bf 26} (2016), no.~3 253--266.

\bibitem{bombelli2009discreteness}
L.~Bombelli, J.~Henson, and R.~D. Sorkin, {\it Discreteness without symmetry
  breaking: a theorem},  {\em Modern Physics Letters A} {\bf 24} (2009), no.~32
  2579--2587.

\bibitem{dowker2020symmetry}
F.~Dowker and R.~D. Sorkin, {\it Symmetry-breaking and zero-one laws},  {\em
  Classical and Quantum Gravity} {\bf 37} (2020), no.~15 155007.

\bibitem{saravani2014causal}
M.~Saravani and S.~Aslanbeigi, {\it On the causal set--continuum
  correspondence},  {\em Classical and Quantum Gravity} {\bf 31} (2014), no.~20
  205013.

\bibitem{rideout1999classical}
D.~P. Rideout and R.~D. Sorkin, {\it Classical sequential growth dynamics for
  causal sets},  {\em Physical Review D} {\bf 61} (1999), no.~2 024002.

\bibitem{zalel2021discrete}
S.~Zalel, {\it Discrete random spacetimes: covariance and quantization in
  growth dynamics for causal sets}, .

\bibitem{benincasa2010scalar}
D.~M. Benincasa and F.~Dowker, {\it Scalar curvature of a causal set},  {\em
  Physical review letters} {\bf 104} (2010), no.~18 181301.

\bibitem{dowker2013causal}
F.~Dowker and L.~Glaser, {\it Causal set d'alembertians for various
  dimensions},  {\em Classical and Quantum Gravity} {\bf 30} (2013), no.~19
  195016.

\bibitem{loomis2017suppression}
S.~Loomis and S.~Carlip, {\it Suppression of non-manifold-like sets in the
  causal set path integral},  {\em Classical and Quantum Gravity} {\bf 35}
  (2017), no.~2 024002.

\bibitem{henneaux1989cosmological}
M.~Henneaux and C.~Teitelboim, {\it The cosmological constant and general
  covariance},  {\em Physics Letters B} {\bf 222} (1989), no.~2 195--199.

\bibitem{buchmuller1989gauge}
W.~Buchm{\"u}ller and N.~Dragon, {\it Gauge fixing and the cosmological
  constant},  {\em Physics Letters B} {\bf 223} (1989), no.~3-4 313--317.

\bibitem{weinberg1989cosmological}
S.~Weinberg, {\it The cosmological constant problem},  {\em Reviews of modern
  physics} {\bf 61} (1989), no.~1 1.

\bibitem{sorkin1997forks}
R.~D. Sorkin, {\it Forks in the road, on the way to quantum gravity},  {\em
  International Journal of Theoretical Physics} {\bf 36} (1997), no.~12
  2759--2781.

\bibitem{BrownYork}
J.~D. Brown and J.~W. York, {\it Jacobi's action and the recovery of time in
  general relativity},  {\em Phys. Rev. D} {\bf 40} (Nov, 1989) 3312--3318.

\bibitem{Unruh:1988in}
W.~G. Unruh, {\it {A Unimodular Theory of Canonical Quantum Gravity}},  {\em
  Phys. Rev. D} {\bf 40} (1989) 1048.

\bibitem{Unruh:1989db}
W.~G. Unruh and R.~M. Wald, {\it {Time and the Interpretation of Canonical
  Quantum Gravity}},  {\em Phys. Rev. D} {\bf 40} (1989) 2598.

\bibitem{Alvarez:2023utn}
E.~Alvarez and E.~Velasco-Aja, {\it {A Primer on Unimodular Gravity}},  1,
  2023.
\newblock \href{http://arxiv.org/abs/2301.07641}{{\tt arXiv:2301.07641}}.

\bibitem{PhysRevD.44.2589}
L.~Bombelli, W.~E. Couch, and R.~J. Torrence, {\it Time as spacetime
  four-volume and the ashtekar variables},  {\em Phys. Rev. D} {\bf 44} (Oct,
  1991) 2589--2592.

\bibitem{Daughton:1998aa}
A.~Daughton, J.~Louko, and R.~D. Sorkin, {\it {Instantons and unitarity in
  quantum cosmology with fixed four volume}},  {\em Phys. Rev. D} {\bf 58}
  (1998) 084008, [\href{http://arxiv.org/abs/gr-qc/9805101}{{\tt
  gr-qc/9805101}}].

\bibitem{herrero2020non}
M.~Herrero-Valea and R.~Santos-Garcia, {\it Non-minimal tinges of unimodular
  gravity},  {\em Journal of High Energy Physics} {\bf 2020} (2020), no.~9
  1--44.

\bibitem{Ellis:2013uxa}
G.~F.~R. Ellis, {\it {The Trace-Free Einstein Equations and inflation}},  {\em
  Gen. Rel. Grav.} {\bf 46} (2014) 1619,
  [\href{http://arxiv.org/abs/1306.3021}{{\tt arXiv:1306.3021}}].

\bibitem{carlip2001quantum}
S.~Carlip, {\it Quantum gravity: a progress report},  {\em Reports on progress
  in physics} {\bf 64} (2001), no.~8 885.

\bibitem{rafael_talk1}
R.~Sorkin, {\it Introduction to everpresent lambda},  nov, 2019.
\newblock PIRSA:19110065 see, \url{https://pirsa.org/19110065}.

\bibitem{kugo2022covariant}
T.~Kugo, R.~Nakayama, and N.~Ohta, {\it Covariant brst quantization of
  unimodular gravity: Formulation with antisymmetric tensor ghosts},  {\em
  Physical Review D} {\bf 105} (2022), no.~8 086006.

\bibitem{hawking1984cosmological}
S.~W. Hawking, {\it The cosmological constant is probably zero},  in {\em
  Euclidean quantum gravity}, pp.~162--163.
\newblock World Scientific, 1984.

\bibitem{coleman}
S.~{Coleman}, {\it {Why there is nothing rather than something: A theory of the
  cosmological constant}},  {\em Nuclear Physics B} {\bf 310} (Dec., 1988)
  643--668.

\bibitem{Ng:1990rw}
Y.~J. Ng and H.~van Dam, {\it {Possible solution to the cosmological constant
  problem}},  {\em Phys. Rev. Lett.} {\bf 65} (1990) 1972--1974.

\bibitem{gielen2022quantum}
S.~Gielen and J.~Magueijo, {\it Quantum resolution of the cosmological
  singularity},  {\em arXiv preprint arXiv:2204.01771} (2022).

\bibitem{Gielen:2013naa}
S.~Gielen, D.~Oriti, and L.~Sindoni, {\it {Homogeneous cosmologies as group
  field theory condensates}},  {\em JHEP} {\bf 06} (2014) 013,
  [\href{http://arxiv.org/abs/1311.1238}{{\tt arXiv:1311.1238}}].

\bibitem{das2023aspects}
S.~Das, A.~Nasiri, and Y.~K. Yazdi, {\it Aspects of everpresent $\lambda$(ii):
  Cosmological tests of current models},  {\em arXiv preprint arXiv:2307.13743}
  (2023).

\bibitem{Feldbrugge:2017kzv}
J.~Feldbrugge, J.-L. Lehners, and N.~Turok, {\it {Lorentzian Quantum
  Cosmology}},  {\em Phys. Rev. D} {\bf 95} (2017), no.~10 103508,
  [\href{http://arxiv.org/abs/1703.02076}{{\tt arXiv:1703.02076}}].

\bibitem{Samuel:2006ga}
J.~Samuel and S.~Sinha, {\it {Surface tension and the cosmological constant}},
  {\em Phys. Rev. Lett.} {\bf 97} (2006) 161302,
  [\href{http://arxiv.org/abs/cond-mat/0603804}{{\tt cond-mat/0603804}}].

\bibitem{Martin:2004xi}
X.~Martin, D.~O'Connor, and R.~D. Sorkin, {\it {The Random walk in generalized
  quantum theory}},  {\em Phys. Rev. D} {\bf 71} (2005) 024029,
  [\href{http://arxiv.org/abs/gr-qc/0403085}{{\tt gr-qc/0403085}}].

\bibitem{Venegas-Andraca:2012zkr}
S.~E. Venegas-Andraca, {\it {Quantum walks: a comprehensive review}},  {\em
  Quant. Inf. Proc.} {\bf 11} (2012) 1015--1106,
  [\href{http://arxiv.org/abs/1201.4780}{{\tt arXiv:1201.4780}}].

\bibitem{Will1}
W.~Cunningham, ``Classical and quantum growth models for discrete spacetime.''
  \url{https://willcunningham.net/trunk/lanl2019.pdf}, 2019.

\bibitem{Will2}
W.~Cunningham, ``Quantum dynamics of total orders.''
  \url{https://willcunningham.net/trunk/odense2020-2.pdf}, 2020.

\bibitem{Padilla:2014yea}
A.~Padilla and I.~D. Saltas, {\it {A note on classical and quantum unimodular
  gravity}},  {\em Eur. Phys. J. C} {\bf 75} (2015), no.~11 561,
  [\href{http://arxiv.org/abs/1409.3573}{{\tt arXiv:1409.3573}}].

\bibitem{Eichhorn:2013xr}
A.~Eichhorn, {\it {On unimodular quantum gravity}},  {\em Class. Quant. Grav.}
  {\bf 30} (2013) 115016, [\href{http://arxiv.org/abs/1301.0879}{{\tt
  arXiv:1301.0879}}].

\bibitem{Wang:2017oiy}
Q.~Wang, Z.~Zhu, and W.~G. Unruh, {\it {How the huge energy of quantum vacuum
  gravitates to drive the slow accelerating expansion of the Universe}},  {\em
  Phys. Rev. D} {\bf 95} (2017), no.~10 103504,
  [\href{http://arxiv.org/abs/1703.00543}{{\tt arXiv:1703.00543}}].

\bibitem{Cree:2018mcx}
S.~S. Cree, T.~M. Davis, T.~C. Ralph, Q.~Wang, Z.~Zhu, and W.~G. Unruh, {\it
  {Can the fluctuations of the quantum vacuum solve the cosmological constant
  problem?}},  {\em Phys. Rev. D} {\bf 98} (2018), no.~6 063506,
  [\href{http://arxiv.org/abs/1805.12293}{{\tt arXiv:1805.12293}}].

\end{thebibliography}\endgroup

\end{document}